\documentclass[]{aa}  

\usepackage{graphicx}
\usepackage{subcaption}
\usepackage{color, epsfig}
\usepackage{apjfonts, natbib}
\usepackage{appendix}
\usepackage{float}
\usepackage{bm}
\usepackage{hyperref}
\usepackage{booktabs}
\usepackage{multirow}
\usepackage{threeparttable}

\begin{document}

   \title{Abnormal Nitrogen Abundance in the X-ray Spectrum of Quasi-periodically Erupting Source AT2019wzc }
   \titlerunning{AT2019wzc N over-abundance}


%
%
%

   \author{Tao Wu\inst{1}, 
           Xinwen Shu\inst{1}\corrauth{xwshu@ahnu.edu.cn},
           Luming Sun\inst{1},
           Ning Jiang\inst{2},
           Jiazheng Zhu\inst{2},
           and Wenjie Zhang\inst{1}
           }
   \authorrunning{T. Wu et al.}
   \institute{Department of Physics, Anhui Normal University, Wuhu, Anhui 241002, People's Republic of China. 
         \and
             School of Astronomy and Space Sciences, University of Science and Technology of China, Hefei, Anhui 230026, People's Republic of China
             }


 
  \abstract
   {Quasi-periodic eruptions (QPEs) are rapid, recurring soft X-ray bursts, whose nature is still in dispute. 
  }
   {A compelling case of QPEs has emerged in the slowly evolving optical transient AT2019wzc, 
possibly associated with the tidal disruption of a post-main-sequence star by a supermassive black hole. Further evidence of a tidal disruption event (TDE) is crucial to understand the nature of AT2019wzc and establish the link between TDE and QPEs.} 
   {We analyzed the high-resolution X-ray spectra of AT2019wzc obtained by XMM-Newton.}
   {We detected a narrow, blueshifted N {\sc vi} absorption line, but weak or undetectable absorption lines from other elements of similar ionization states such as carbon and oxygen. 
    The absorption line features can be reproduced by an ionized gas with ionization parameter $\log \xi \sim 0.3\ {\rm erg~cm~s^{-1}}$ and column density $N_{\rm H}\sim 10^{20}\ {\rm cm^{-2}}$, under the condition of a nitrogen abundance of $11.6_{-7.8}^{+19.6}$ times the solar value.}
   {This abnormal nitrogen abundance favors a TDE origin for AT2019wzc, and the absorbing gas may originate from the outflow induced by self-collision of the TDE's debris stream.}


   \maketitle
   \nolinenumbers

\section{Introduction} \label{sec:Introduction}

X-ray quasi-periodic eruptions (QPEs) are repeating, high-amplitude, soft X-ray bursts observed from the nuclei of low-mass galaxies. 
The QPE candidates were first proposed in \cite{Sun2013} and then confirmed by \cite{Miniutti2019} in the QPE source GSN 069. 
Since the first discovery, 
the number of known QPE sources has grown to more than a dozen \citep{Miniutti2019, Giustini2020,Arcodia2021,Arcodia2024,Arcodia2025,Chakraborty2021,Chakraborty2025a,Quintin2023,Nicholl2024,Sa'nchez-Sa'ez2024,Hernandez-Garcia2025}. 
Despite this growing sample, the physical origin of QPEs remains an open question in the study of X-ray variability of nuclear transients. 

One of intriguing scenarios to explain QPEs involves a stellar-mass object orbiting a supermassive black hole (SMBH) in an extreme mass-ratio inspiral (EMRI) configuration.
As a star or a stellar-mass compact object crosses a pre-existing accretion disk, it produces two flares per orbit \citep{Dai2010,Sukova2021,King2020,Xian2021,Wang2022,King2022,Zhao2022,Lu2023,Franchini2023,Linial2023b,Tagawa2023,Kejriwal2024,Zhou2024a}.
In some systems, X-ray or optical tidal disruption events (TDEs, \citealt{Rees1988}) were observed prior to the onset of QPEs \citep{Shu2018, Miniutti2019,  Chakraborty2021,Nicholl2024,Bykov2025,Chakraborty2025a}.
This apparent TDE–QPE connection has motivated a unified model, in which QPEs represent a transient phase following a TDE, occurring shortly after the cessation of AGN activity \citep{Jiang2025,Linial2023b}. 
However, it remains unclear whether this model applies to all QPEs and whether a TDE is a necessary precursor for QPE production. 
Therefore, 
identifying TDE features in QPEs is essential for testing the broad applicability of this model.


Recent studies suggest that enhanced nitrogen (N) abundance could serve as a novel diagnostic for identifying and confirming TDE candidates, particularly for transient sources with otherwise ambiguous origins \citep{Cenko2016,Kochanek2016,Yang2017,Miller2023,Sheng2021,Mockler2022,Mockler2024,Kosec2025}. 
Observationally, enhanced N abundances have been detected in the 
UV and X-ray spectra of several TDEs, such as ASASSN-14li and GSN 069 \citep{Cenko2016,Yang2017,Miller2023,Sheng2021,Kosec2025}, and these signatures differ markedly from those seen in typical AGNs.
Stellar evolution analysis indicates that this N enrichment can be naturally explained by the tidal disruption of a near-solar-mass main-sequence star \citep{Kochanek2016}, as it releases internal material whose nitrogen abundance has been enhanced by the CNO cycle \citep{Iben1964,Iben1967,Lambert1977,Lambert1981}.
Consequently, enhanced N abundance not only marks a TDE but also indicates the tidal disruption of a moderately massive star  \citep{Mockler2022,Miller2023}.

Among the known QPE sources, AT2019wzc, a system hosting a $10^6~M_\odot$ SMBH at redshift $z=0.024$, is particularly interesting  due to its extreme timing properties, including the longest durations and recurrence intervals \citep{Hernandez-Garcia2025}.
However, the physical origin of its QPEs remains ambiguous.
Multi-wavelength observations suggest two competing scenarios: a turn-on AGN or a TDE \citep{Sa'nchez-Sa'ez2024,Hernandez-Garcia2025}.
More recently, \citet{Zhu2025} analyzed its UV spectrum and proposed an association with a featureless TDE \citep{Hammerstein2023,Yao2025,Anna2025}, based on the lack of UV emission lines and the extremely steep UV spectral slope.
{Nevertheless, these indications do not conclusively establish a TDE origin, and more direct evidence is required.}



In this work, we analyze the archival 
high-resolution X-ray spectra of AT2019wzc obtained by {\it XMM-Newton},   
and report the detection of an isolated N {\sc vi} absorption line, which indicates an enhanced N abundance.
This discovery provides further evidence of linking the QPEs in AT2019wzc to a TDE.
The paper is structured as follows.
Section~\ref{sec:Reduction} describes the {\it XMM-Newton} reflection grating gpectrometer (RGS) data reduction.
Section~\ref{sec:results} presents the spectral analysis results.
Section~\ref{sec:discussion} discusses the implications of nitrogen enhancement for a TDE origin. 
Section~\ref{sec:conclusion} summarizes our conclusions.

\begin{figure*}
\centering
  \includegraphics[width=\linewidth]{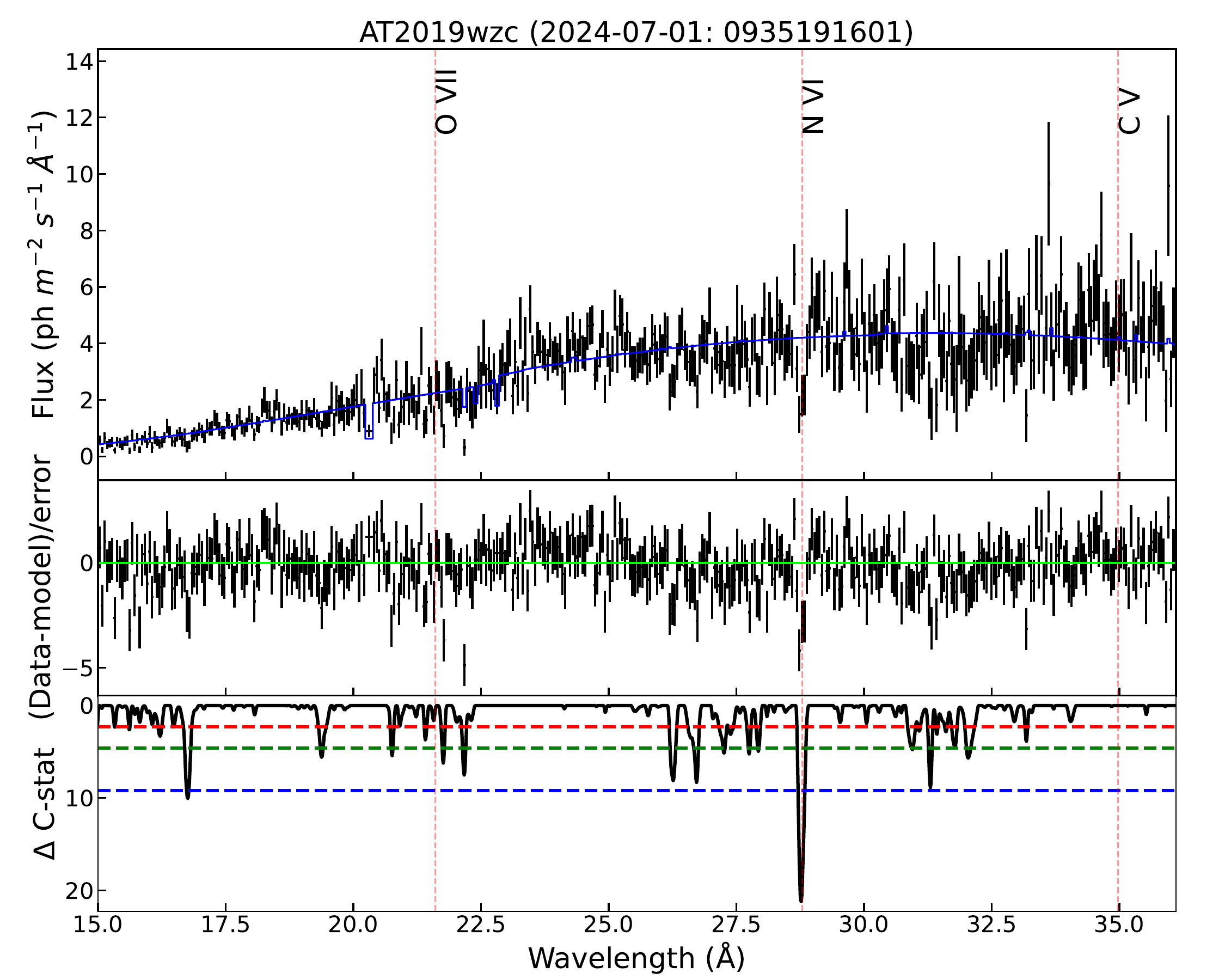}
\caption{{Absorption line search in the XMM2/RGS spectrum of AT2019wzc.}
\textbf{Top panel:} RGS spectrum (black) obtained on 2024 July 1. 
The best-fitting continuum model is shown in blue.
\textbf{Middle panel:} 
A significant deficit is visible at 28.77~\AA. 
\textbf{Bottom panel:} Results of the Gaussian line scan. The $y$-axis shows the improvement in $\Delta C$ upon adding a Gaussian absorption line with a fixed width of 0.02~\AA. 
Red vertical dashed lines mark the rest-frame wavelengths of C {\sc v} (34.9728~\AA), N {\sc vi} (28.7875~\AA), and O {\sc vii} (21.6019~\AA). 
The N {\sc vi} absorption line is prominent, while no significant features are detected at the positions of C {\sc v} and O {\sc vii}.
\label{fig:blind_search}
}
\end{figure*}

\section{Observations and data reduction} \label{sec:Reduction} 

This paper focuses on two {\it XMM-Newton} \citep{Jansen2001} observations of AT2019wzc, taken on 27 June 2024 (XMM1 hereafter, ObsID 0935191401) and 1 July 2024 (XMM2, ObsID 0935191601).
XMM1 captured the declining phase of a QPE, while XMM2 was conducted just after the peak of the subsequent QPE.
For both observations, the RGS data \citep{denHerder2001} that have best spectroscopy resolution are available.
Note that we did not use the archival observation taken on 29 June 2024 (ObsID 0935191501), as it was conducted during quiescence and lacks RGS observing mode. 

We retrived the {\it XMM-Newton} data from the XMM-Newton Science Archive (XSA).
The RGS data was reduced using the standard Science Analysis System (SAS, version 21.0.0) with the corresponding calibration files. 
We generated the source and background spectra, and response matrix using the {\it rgsproc} task, and then converted them from the \textsc{OGIP} format to the {\it SPEX} (v. 3.08.01) format \citep{Kaastra1996} using the {\it trafo} routine.
To optimize spectral sensitivity, we restricted our analysis to the combined first-order spectra of RGS 1 and RGS 2 over the 15–37~\AA\ band.
The total RGS counts are 4506 for XMM1 and 10,456 for XMM2.


We binned the RGS spectra by a factor of 5 to achieve mild oversampling relative to the instrumental spectral resolution, and adopted the Cash statistic (Cash 1979).
Unless otherwise stated, all quoted statistical errors correspond to the 68\% confidence level ($\Delta C = 1$), and the upper/lower limits correspond to the 90\% confidence level ($\Delta C = 2.7$).


\begin{table}
\caption{Spectral fits of the ionized absorption using the {\it pion} model \label{tab:pion}}
\begin{tabular}{c|cc}
\hline
\hline
Component & Parameter & XMM2 \\
\hline
blackbody & kT (eV) & $76.7_{-1.1}^{+1.1}$\\
 & {\it C/dof} & 503.3/433\\
\hline
line & $wl$ (\AA) & $28.77_{-0.02}^{+0.02}$\\
 & $awhg$ (\AA) & $0.077_{-0.057}^{+0.036}$\\
 & $\Delta C$ & 21.5\\
\hline
pion & $N_{\rm H}$ ($10^{22}~cm^{-2}$) & $0.0037_{-0.0032}^{+0.0054}$\\
 & $\log \xi$ (${\rm erg~cm~s}^{-1}$)& $0.3_{-0.3}^{+0.4}$\\
 & turbulent velocity (${\rm km~s}^{-1}$) & $401_{-144}^{+168}$\\
 & blueshifted velocity (${\rm km~s}^{-1}$) & $-102_{-194}^{+206}$\\
 & $A_{\rm N}$ & $11.6_{-7.8}^{+19.6}$\\
 & $\Delta C$ & 17.4\\
\hline
\end{tabular}
\begin{tablenotes}
  \item Note: XMM2/RGS spectral fitting results obtained with {\it SPEX}.
  Ionized absorption associated with the central engine was modeled using the {\it pion} photoionization component.
  Both the blackbody emission and ionized absorption were corrected for the host galaxy redshift. Therefore, the reported blueshifted velocity is relative to the host galaxy rest frame.
  The nitrogen abundance was treated as a free parameter. 
\end{tablenotes}
\label{tab:pion}
\end{table}

\section{Analysis and results} \label{sec:results}

\subsection{The isolated N {\sc vi} absorption line in XMM2/RGS} \label{subsec:Blind search}

As reported in previous studies \citep{Sa'nchez-Sa'ez2024, Hernandez-Garcia2025, Chakraborty2025b}, the XMM2/RGS spectra of AT2019wzc are super-soft and well described by a blackbody. 
For preliminary continuum modeling, we adopted the same baseline model as in the literature: a simple blackbody ({\it bb}) component, modified by nearly neutral gas absorption in both the host galaxy and the Milky Way, modeled with {\it absm}. 
The host galaxy column density was left free, while the Milky Way absorption was fixed at $6.4 \times 10^{20}\ {\rm cm^{-2}}$. 
We also included the {\it reds} model to account for the redshift of AT2019wzc ($z=0.024$, \citealt{Sa'nchez-Sa'ez2024,Hernandez-Garcia2025}).

After the continuum modeling, we performed a blind search for absorption lines, which are sensitive to outflow properties. 
In this procedure, we added a Gaussian absorption line to the continuum model with the optical depth set as a free parameter.
We tried central wavelengths on a grid spanning 15–37~\AA\ with a step size of 0.01~\AA.
The line width was fixed at 0.02~\AA during the blind search. Note that we also explored other line widths (0.04, 0.06, and 0.08~\AA) and found that the choice of width has a negligible impact on the line detection.
At each grid point, we refitted the spectrum with the additional line component and recorded the $\Delta C$ statistic relative to the baseline continuum model. 
We adopted a $\Delta C$ threshold of 10 for line search, corresponding to a confidence level of $>99$\% for two free parameters, but no absorption lines were detected in XMM1 (see Section~\ref{subsec:XMM1_no_line}).
In contrast, the line scan revealed a prominent absorption feature in XMM2 with $\Delta C \gtrsim 20$ at 28.77~\AA, as shown in Fig.~\ref{fig:blind_search}, but no other significant absorption lines present. 
The continuum parameters for XMM2 are summarized in Table~\ref{tab:pion}. 
This baseline model yields C statistics of 503.3 with degrees of freedom $\nu = 433$. 
The blackbody temperature is $kT = 76.7_{-1.1}^{+1.1}$~eV, consistent with the values reported by \citet{Chakraborty2025b}. 


To further characterize the absorption feature in XMM2/RGS, we fitted it with a Gaussian profile ({\it line}), allowing for both the wavelength and width to vary freely. 
The results are summarized in Table~\ref{tab:pion} and Fig.~\ref{fig:blind_search}, yielding a centroid wavelength of $28.77_{-0.02}^{+0.02}$~\AA, a lower limit of optical depth $\tau > 0.9$, and a line width $awhg = 0.077_{-0.057}^{+0.036}$~\AA. 
The equivalent width (EWs) inferred from the model is $0.11_{-0.02}^{+0.02}$~\AA. 
The line width (awhg) indicates that the velocity width of the absorber is $\sim 800~{\rm km~s}^{-1}$. 
In this case, the improvement in the C-statistic is $\Delta C = 21.5$, corresponding to a detection significance of approximately $4\sigma$ assuming Gaussian distribution. 
In Fig.~\ref{fig:blind_search}, we mark the rest-frame wavelengths of transitions from strong lines of interest in the range of XMM2/RGS spectrum, such as C {\sc v} (34.9728~\AA), N {\sc vi} (28.7875~\AA), and O {\sc vii} (21.6019~\AA). 
Based on the wavelength coincidence, we identified the absorption feature at $28.77_{-0.02}^{+0.02}$~\AA~as N {\sc vi}, although further verification through photoionization modeling is necessary.


\begin{figure*}
\centering
  \includegraphics[width=\linewidth]{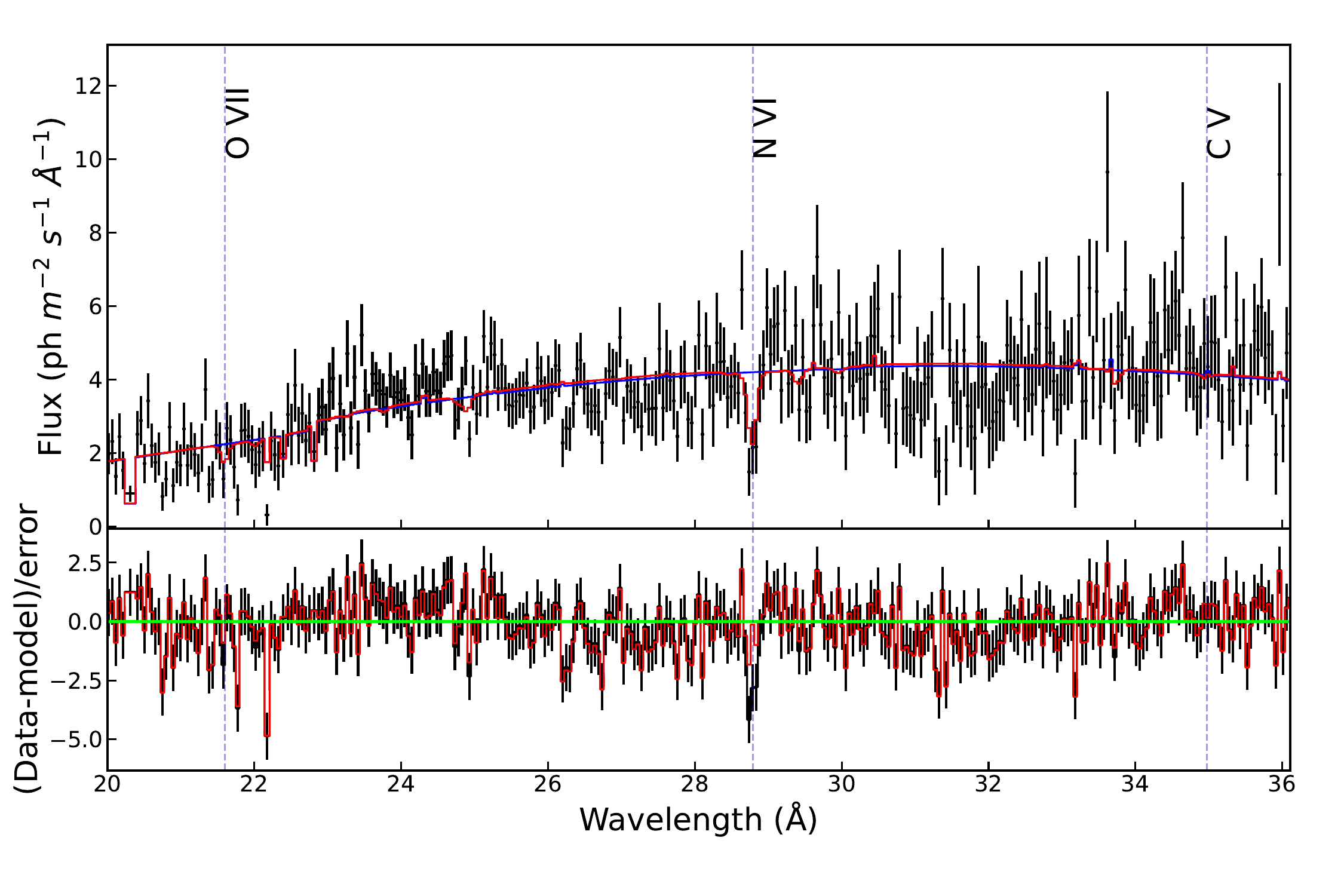}

\caption{{Photoionization modeling of the absorber in the XMM2/RGS spectrum.}
\textbf{Top panel:} The best-fitting continuum-only model is shown in blue, while the photoionization model ({\it pion}) with enhanced nitrogen abundance is shown in red.
\textbf{Bottom panel:} The significant residuals at 28.77~\AA~are well described by the absorber with an N enhancement. The positions of the most relevant transitions (C {\sc v}, N {\sc vi}, and O {\sc vii}) are indicated by vertical dashed lines.
\label{fig:pion}
}
\end{figure*}

\subsection{Enhanced N abundance in the absorber} \label{subsec:pion}

The N {\sc vi} line is not typically the strongest absorption feature in AGN or TDE spectra.
To understand the physical conditions required to produce a strong, isolated N {\sc vi} line in AT2019wzc, 
we simulated the absorption-spectrum model to fit the data using the photoionization code {\it pion} in {\it SPEX} \citep{Miller2023}. 
Initially, we assumed a solar abundance for the absorber, and set the column density ($N_{\rm H}$), ionization parameter ($\log \xi$), turbulent velocity ($V_{\rm RMS}$), and blueshifted velocity ($V_{\rm out}$) as free parameters. 
After detailed fittings, the {\it pion} model identified the 28.77~\AA\ feature as N {\sc vi}. However, the model with solar abundances could not fully reproduce the observed line profile, underpredicting the EW of the N {\sc vi} line by a factor of $\sim$3. This discrepancy suggests that an enhanced nitrogen abundance is required.


Previous analysis of the RGS spectrum of the TDE ASASSN-14li  \citep{Miller2023} revealed a strong N over-abundance ($> 100$ times solar value), suggesting that a similar scenario could apply to AT2019wzc.
We then allowed the N abundance to vary freely in the {\it pion} model and refitted the spectrum.
The fit results are summarized in Table~\ref{tab:pion}. 
This model improves the C-statistic by $\Delta C = 17.4$ relative to the continuum model.  
As shown in Fig.~\ref{fig:pion}, the {\it pion} model with N abundance enhanced by a factor of $A_{\rm N} = 11.6_{-7.8}^{+19.6}$ successfully reproduces the strongest, isolated absorption line at 28.77~\AA, without generating any other detectable strong absorption lines.
Therefore, the XMM2 spectrum can be reasonably explained by the absorber with a higher N abundance.



\begin{figure*}
    \centering
\begin{subfigure}[b]{0.45\textwidth}
    \centering
    \includegraphics[width=\textwidth]{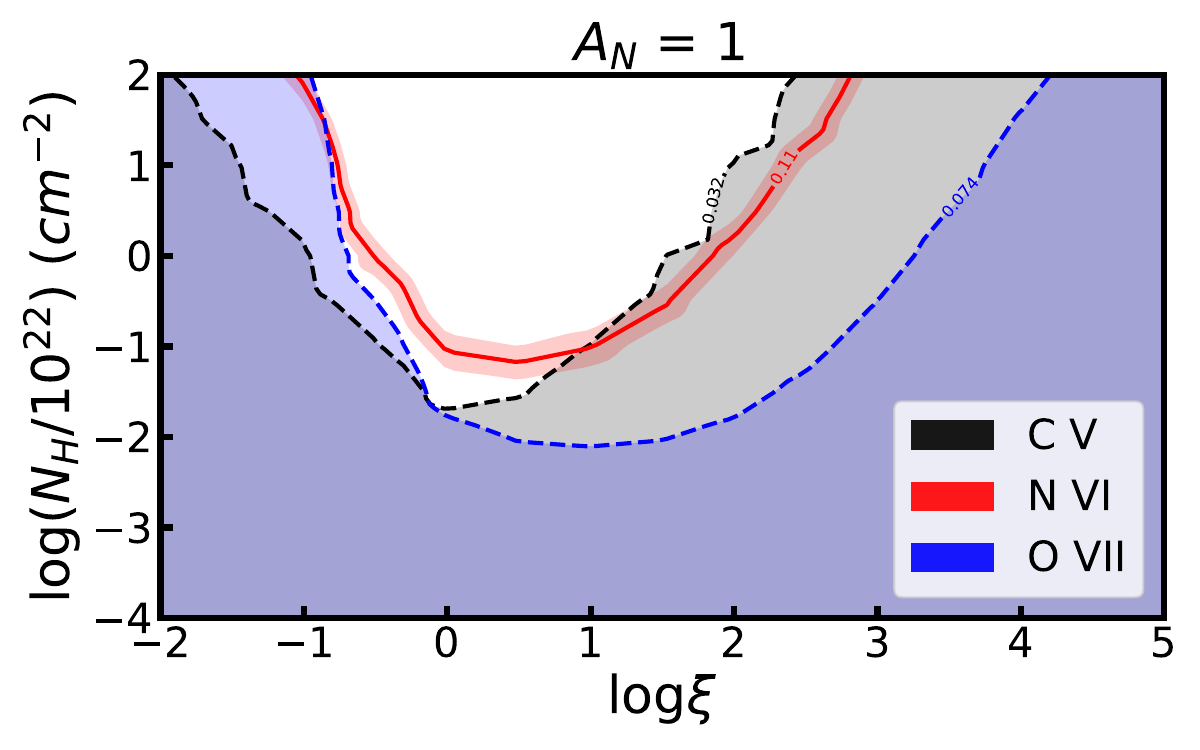}
    \caption{}
\end{subfigure}
\hfill
\begin{subfigure}[b]{0.45\textwidth}
    \centering
    \includegraphics[width=\textwidth]{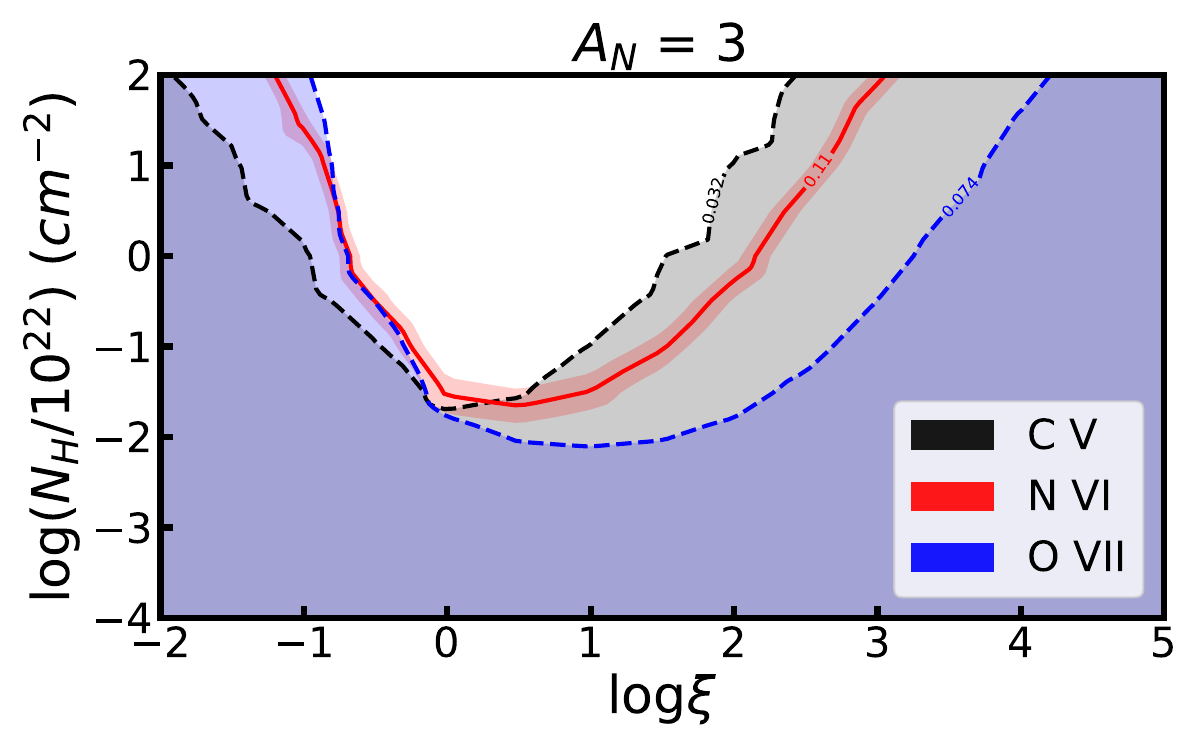}
    \caption{}
\end{subfigure}
\begin{subfigure}[b]{0.45\textwidth}
    \centering
    \includegraphics[width=\textwidth]{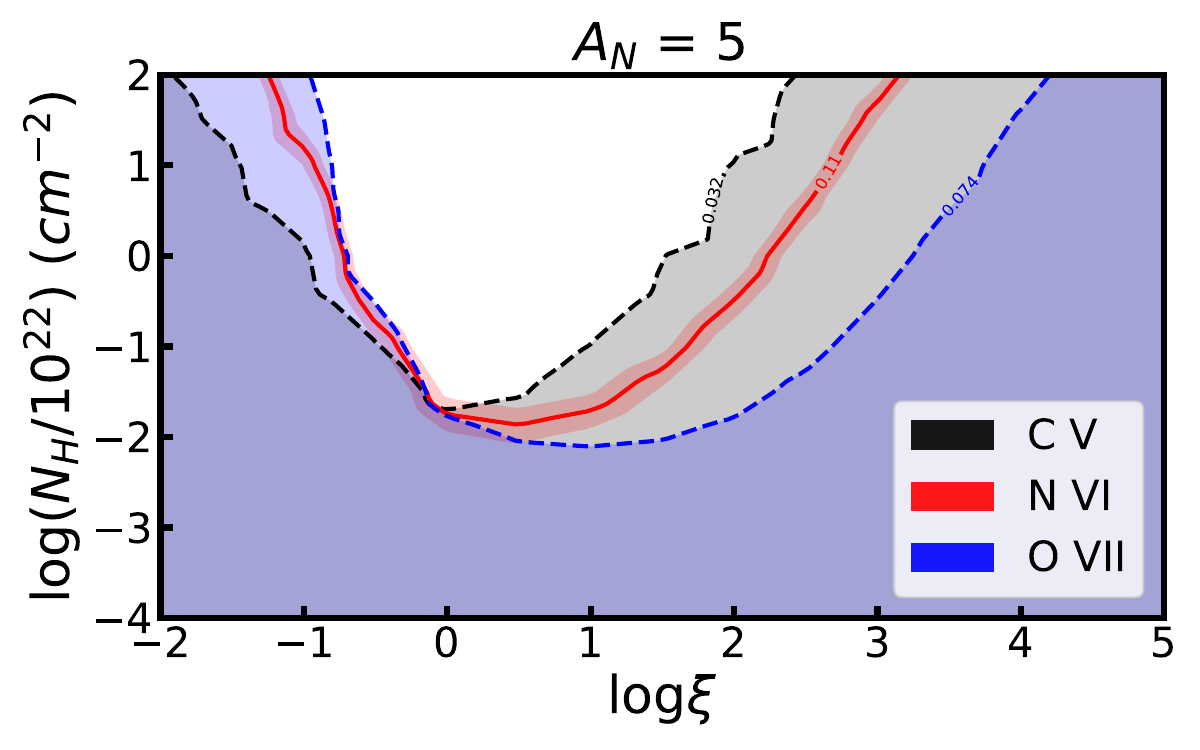}
    \caption{}
\end{subfigure}
\hfill
\begin{subfigure}[b]{0.45\textwidth}
    \centering
    \includegraphics[width=\textwidth]{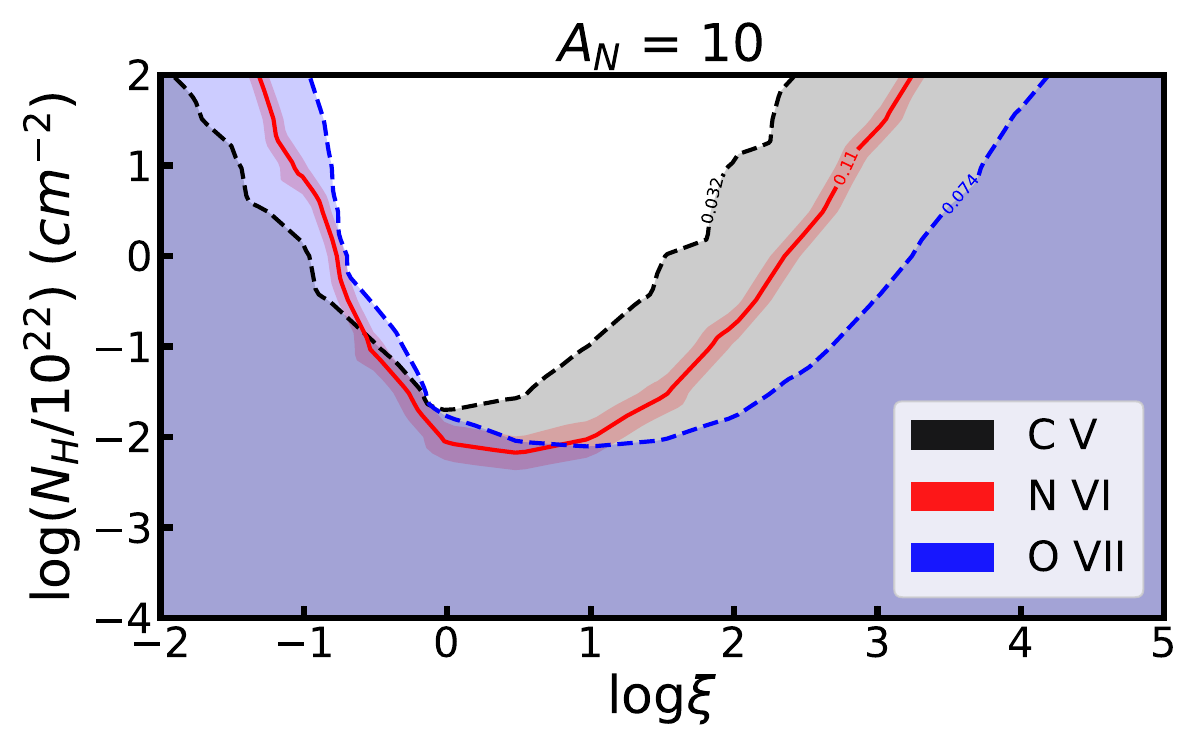}
    \caption{}
\end{subfigure}
\caption{
Parameter space for producing an isolated N {\sc vi} absorption line.
The red region represents the 68\% confidence interval of the observed EW of N {\sc vi} (28.7875~\AA).
The black and blue correspond to the constraints from C {\sc v} (34.9728~\AA) and O {\sc vii} (21.6019~\AA), respectively, with dashed lines indicating their 90\% upper limits.
Panels (a)–(d) show models with increasing nitrogen abundance from 1 to 10 times solar.
An overlapping region that satisfies all observational constraints emerges only when $A_N\gtrsim3-5$. 
\label{fig:pion_cal}
}
\end{figure*}

\subsection{Physical conditions of ionized gas} \label{subsec:physical conditions}

We now explore which physical conditions can produce the N {\sc vi} while suppressing other strong absorption lines within the spectral range of RGS.
The best-fitting {\it pion} model gives an ionization parameter $\log \xi \sim 0.3\ {\rm erg~cm~s^{-1}}$, which ensures that nitrogen dominantly exists as N {\sc vi} (ionization potential of 97.9 eV) rather than N {\sc v} (77.5 eV) or N {\sc vii} (552 eV).
This favors the production of the N {\sc vi} He$\alpha$ line at 28.787~\AA\ while suppressing the N {\sc vii} Ly$\alpha$ line at 24.785~\AA.
The column density $N_{\rm H} \sim 0.004 \times 10^{22}\ {\rm cm^{-2}}$ is appropriate for producing the observed N {\sc vi} optical depth ($\tau~>~0.9$).
Compared to typical warm absorbers (WA) in AGNs, which have $\log \xi \sim 0$–3 and $N_{\rm H} \sim 10^{20}$–$10^{22}\ {\rm cm^{-2}}$ \citep{Tombesi2013}, the inferred physical conditions of the absorber in AT2019wzc exhibit lower values for both parameters.

Under these conditions, carbon mainly exists as C {\sc v} (64.5 eV), while oxygen can be found in O {\sc v} (77.4 eV), O {\sc vi} (113.9 eV), and O {\sc vii} (138.1 eV).
Within the observed 15–37~\AA\ band, the strong absorption lines from these ionized states include the C {\sc v} He$\beta$ line at 34.973~\AA\ and the O {\sc vii} He$\alpha$ line at 21.602~\AA.
Neither C {\sc v} nor O {\sc vii} is significantly detected in the spectrum.
To estimate upper limits on their EWs, we added Gaussian lines with fixed central wavelengths and widths matching that of N {\sc vi}.
The resulting 90\% confidence upper limits for C {\sc v} and O {\sc vii} are 0.032~\AA\ and 0.074~\AA, respectively. 

Using the XMM2/RGS continuum model as a baseline, we computed a grid of {\it pion} models by varying $N_{\rm H}$, $\log \xi$, and $A_{\rm N}$. 
The column density and ionization parameter were varied in steps of 0.5~dex, while the nitrogen abundance was varied in steps of 1 solar abundance. 
In all calculations, the turbulent velocity and blueshifted velocity were fixed at 400~km~s$^{-1}$ and 0~km~s$^{-1}$, respectively.
We explored the $N_{\rm H}$–$\log \xi$ parameter space and identified regions where the predicted EWs of N {\sc vi}, C {\sc v} and O {\sc vii} match the observed values. 
Fig.~\ref{fig:pion_cal} shows four slices of this space with N abundances of 1, 3, 5, and 10 times solar.
In each panel, the red belt indicates the 68\% confidence interval of the observed N {\sc vi} EW, while black and blue shades represent the 90\% upper limits for C {\sc v} and O {\sc vii}, respectively. 
Only when the N abundance $A_N\gtrsim3-5$ does a narrow parameter space ($N_{\rm H} \sim 10^{20}\ {\rm cm^{-2}}$, $\log \xi \sim 0$) emerge that can simultaneously reproduce all three lines.
Higher N abundances expand this allowed parameter space, establishing a lower limit of $A_{\rm N} > 3$. 

Our calculations demonstrate that producing an isolated N {\sc vi} absorption line is theoretically feasible but requires rather extreme physical conditions: elevated N abundance ($A_N\gtrsim3-5$) and a restricted range of column density and ionization parameter ($N_{\rm H} \sim 10^{20}\ {\rm cm^{-2}}$, $\log \xi \sim 0$). 
We will continue the discussion in Section~\ref{subsec:TDE origin}. 


\begin{table*}
\caption{Comparison of the parameters of low-speed absorbers with other  TDEs.}
\begin{tabular}{c|ccccccc}
\hline
\hline
Name & Ref. & $M_{\rm BH}$ & $N_{\rm H}$ & $\log \xi$ & $v_{\rm turb}$ & $v_{\rm out}$ & $A_{\rm N}$  \\
  &  & $M_{\odot}$ & $10^{22}~{\rm cm}^{-2}$ & ${\rm erg~cm~s}^{-1}$ & ${\rm km~s}^{-1}$ & ${\rm km~s}^{-1}$ & \\
 (1) & (2) & (3) & (4) & (5) & (6) & (7) & (8) \\
\hline
AT2019wzc & This work & $\sim~10^{6}$ & $0.0037_{-0.0032}^{+0.0054}$ & $0.3_{-0.3}^{+0.4}$ & $401_{-144}^{+168}$ & $102_{-194}^{+206}$ & $11.6_{-7.8}^{+19.6}$ \\
ASASSN-14li & \citet{Miller2015, Miller2023} & $\sim~10^{6}$ & $0.1-1.3$ & $3.6-4.1$ & $60-230$ & $130-500$ & $109-160$ \\
ASASSN-20qc & \citet{Kosec2023} & $\sim~2~\times~10^{7}$ & $0.13-0.21$ & $1.0-1.9$ & $240-270$ & $380-410$ & -- \\
 & & & $7-24$ & $3.7-3.8$ & $80-90$ & $910$ & -- \\
GSN 069 & \citet{Kosec2025} & $\sim~10^{6}$ & $1.0-1.7$ & $3.9-4.6$ & $400-1100$ & $1700-2900$ & $11-50$ \\
\hline
\end{tabular}
\begin{tablenotes}
  \item (1) source name; (2) references; (3) BH mass; (4) column density; (5) ionization parameter; (6) turbulent velocity; (7) blueshifted velocity; (8) N over-abundance relative to the solar one.
\end{tablenotes}
\label{tab:Comparison}
\end{table*}

\section{Discussion} \label{sec:discussion}

\subsection{Absence of the N {\sc vi} absorption line in XMM1} \label{subsec:XMM1_no_line}

While we detected a significant, isolated N {\sc vi} absorption line in the XMM2/RGS spectrum, it is absent in XMM1 observation taken 4 days prior to the XMM2 observation. 
The blueshifted velocity and width of the absorption line are only several $10^2$ ${\rm km\ s^{-1}}$, corresponding to a long dynamic timescale of the absorber, which indicates that it is unlikely to appear or move into the line of sight within just 4 days.
There are two possible interpretations for the non-detection of N {\sc vi} in XMM1: either the N {\sc vi} EW decreased as the ionization parameters $\log \xi$ changed, or the poorer data quality of XMM1 reduced the signal-to-noise ratio (SNR) of the absorption line. 
The former is based on the fact that the X-ray luminosity in the range of 15–37 \AA\ for XMM1 ($L_{\rm X}=4.5 \times 10^{42}~{\rm erg~s^{-1}}$) is a factor of $>2$ lower than that in XMM2 ($L_{\rm X}=9.3 \times 10^{42}~{\rm erg~s^{-1}}$).

To test these possibilities, we performed extensive spectral simulations. 
In the simulations, we used the continuum model and exposure time of XMM1.
Most of the parameters to characterize the absorber were adopted from the results obtained in Section~\ref{subsec:pion}, except for $\log \xi$, which was decreased to match the lower luminosity observed by XMM1.
We then fitted the N {\sc vi} absorption feature in the simulated spectrum. 
The resulting line is nearly undetectable, with $\Delta C \lesssim 6$. 
This confirms that reduced luminosity and lower SNR can indeed make the N {\sc vi} absorption line undetectable. 
At the same time, this result suggests that the enhanced nitrogen abundance observed in AT2019wzc appears not to be a transient phenomenon, but may instead persist over time. 
Future observations with higher spectral quality will be required to verify this possibility. 



\subsection{Enhanced nitrogen abundance supports a TDE origin} \label{subsec:TDE origin}

The nature of AT2019wzc is under debate in the literature.
Previous works that reported the discovery of AT2019wzc \citep{Sa'nchez-Sa'ez2024,Hernandez-Garcia2025} noticed evidence supporting a TDE scenario, such as the soft X-ray spectrum and the absence of broad emission lines.
Nevertheless, a turn-on AGN scenario was still favored, because the overly long rise and decay timescales in the light curve, as well as the [OIII] emission line echo, are characteristics that 
are different from most of the known TDEs.
On the contrary, \cite{Zhu2025} recently interpreted the long evolution timescales with a TDE of a post-main-sequence star ($M_\star \sim 1~M_{\odot}$, $R_\star \sim 3$–$10~R_{\odot}$), which can also explain the lower peak luminosity of AT2019wzc in comparison with other TDEs. 
Furthermore, they found an extremely steep UV-to-optical SED with a powerlaw index of $-2.6$, which does not match the prediction from the AGN accretion disk theory but supports the TDE scenario.
Due to the absence of broad emission lines in the UV and optical bands, they classified AT2019wzc as a featureless TDE \citep{Hammerstein2023,Yao2025,Anna2025}.



The photoionization modeling of the XMM2/RGS spectrum presented in section~\ref{subsec:pion} yields a significant N over-abundance, with a best-fit value of $A_N\sim11$ and a conservative lower limit of $A_N\gtrsim3-5$. 
An independent analysis of the same data found a N over-abundance of $A_N\sim22$ in a much faster ($\sim0.04c$) outflow emission component \citep{Chakraborty2025b}, confirming the presence of N-rich gas across different kinematic regimes. 
Similar N enrichment with $A_N\sim10-100$ has been observed in multiple TDEs across UV and X-ray bands, in both emission and absorption spectra \citep{Cenko2016, Yang2017, Sheng2021, Miller2023, Kosec2025}.
Nitrogen over-abundance has also been observed in a small fraction of AGNs \citep[e.g., ][]{Baldwin2003, JiangLinhua2008, Isobe2025}.
However, the N enhancement in these AGNs is generally modest, typically $\lesssim10$ relative to solar.
Furthermore, its origin is either attributed to tidal disruption events that occasionally contaminate the AGN broad-line region \citep{LiuXin2018}, or to star formation and chemical enrichment within the accretion disk or nuclear region \citep{HuangJiamu2023}, which requires prolonged star formation activity over extended timescales.
Neither scenario is compatible with the turn-on AGN hypothesis for AT2019wzc, as there is no evidence of long-term AGN or star formation activities.
Thus, the enhanced nitrogen abundance observed in AT2019wzc provides independent evidence that is more naturally explained by a TDE origin than by the turn-on AGN scenario.



\subsection{The possible physical origin of low-velocity absorber} \label{subsec:TDE origin}

The absorber has a line-of-sight velocity of $-102^{+206}_{-102}$ ${\rm km~s}^{-1}$.
Its line width ($awhg=0.077_{-0.057}^{+0.036}$ \AA) is close to the RGS spectral resolution of 0.06--0.07 \AA, implying an intrinsic width of $\lesssim700$ ${\rm km~s}^{-1}$, consistent with the turbulence velocity of $401_{-144}^{+168}$ ${\rm km~s}^{-1}$ derived from the \textit{pion} fittings.
Similar low-velocity absorbers have been detected in ASASSN-14li ($v\sim100-500$ ${\rm km~s}^{-1}$, \citealt{Miller2015,Miller2023}) and ASASSN-20qc ($v\sim900$ ${\rm km~s}^{-1}$, \citealt{Kosec2023}), as summarized in Table~\ref{tab:Comparison}.
The detection of such narrow absorption lines in multiple TDEs indicates that the absorbing gas covers a substantial solid angle as seen from the central source.
\cite{Miller2015} considered that the absorber in ASASSN-14li could originate from stellar debris near apocenter, but the small covering factor would result in a low probability of intercepting the line of sight, making this interpretation less favorable.


Alternatively, the narrow, low-velocity absorbers in these TDEs could be  explained by the collision-induced outflow (CIO) from the self-intersection of the debris stream \citep{Lu_Bonnerot2020}.
For AT2019wzc and ASASSN-14li with black hole mass of $\sim10^6\ M_\odot$, the relativistic apsidal precession is relatively weak, placing the stream self-intersection at large radii.
According to the simulations of \cite{Lu_Bonnerot2020}, the CIO launched from such an intersection has an initial velocity of $\sim0.01c$.
The outermost regions of CIO are expected to have lower velocities, and the observed line-of-sight velocity of the absorber can be further reduced by gravitational deceleration or projection effects, naturally accounting for the low observed velocities (hundreds ${\rm km~s}^{-1}$).
For ASASSN-20qc, which hosts a more massive black hole ($\sim2\times10^7 M_\odot$), the predicted CIO initial velocity is higher, consistent with the two observed components with larger velocities ($400-900$ ${\rm km~s}^{-1}$).
This CIO picture is further supported if the TDE in AT2019wzc originated from a post-main-sequence star \citep{Zhu2025}.
The larger stellar radius would place the stream self-intersection at a greater distance, naturally producing the lower velocity, lower ionization parameter and lower column density observed in AT2019wzc compared to the other sources (Table~\ref{tab:Comparison}).

In contrast to the narrow, low-velocity absorbers, broad and transient ultrafast outflows (UFOs) with velocities of $\gtrsim0.1c$ have been detected in ASASSN-14li \citep{Kara2018} and AT2019wzc \citep{Chakraborty2025b}.
These fast outflows can be attributed to super-Eddington accretion disk winds \citep[e.g., ][]{Coughlin2014, McKinney2015}.
GSN 069 exhibits an outflow with velocities of $\sim1700-2900$ ${\rm km~s}^{-1}$ \citep{Kosec2025}, which lies between the slow CIO components and the UFOs discussed above.
Its origin is therefore less certain.
However, GSN 069 has a smaller black hole mass, which yields lower orbital velocities at the site of stream intersection. In addition, its outflow has a long evolutionary timescale of about 20 yr \citet{Miniutti2019}, during which it could have decelerated. Both factors would tend to produce a CIO with lower velocities than observed. 
Therefore, the CIO scenario appears disfavored in GSN 069. 

\subsection{Future improved constraints with HUBS}
\label{HUBS}

While it is sufficient 
to detect and characterize the N {\sc vi} absorption line in AT2019wzc, the current {\it XMM-Newton}/RGS data are not sensitive enough to place a strong constraint on the nitrogen abundance, due to the lack of detections of absorption lines from C and O ions. 
The Hot Universe Baryon Surveyor (HUBS, \citealt{Bregman2023}) will offer 
a spectral resolution of $\sim2$ eV and an effective area several times larger than that of RGS in the soft X-ray band, which is important to improve the 
constraints on the nitrogen abundance as well as the kinematic properties of the absorber. 

\begin{figure*}
    \centering
\begin{subfigure}[b]{0.96\textwidth}
    \centering
    \includegraphics[width=\textwidth]{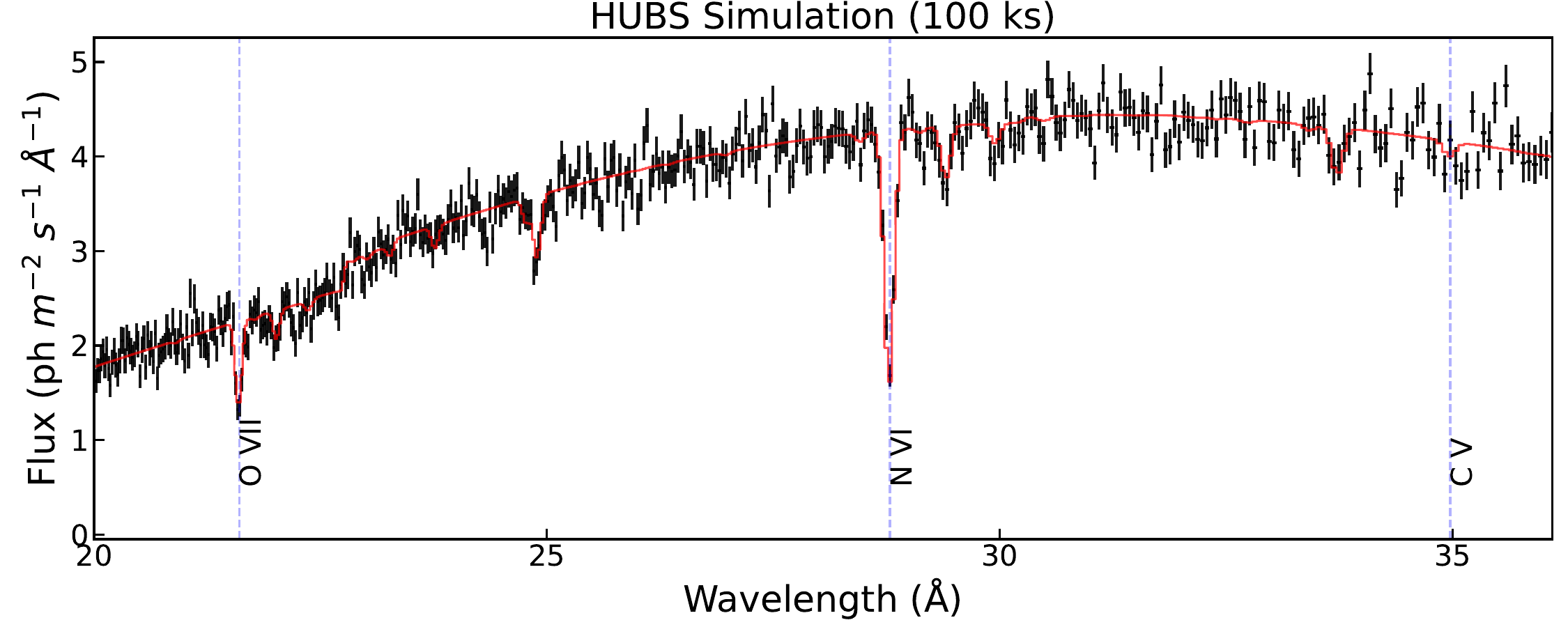}
    \caption{}
\end{subfigure}
\begin{subfigure}[b]{0.45\textwidth}
    \centering
    \includegraphics[width=\textwidth]{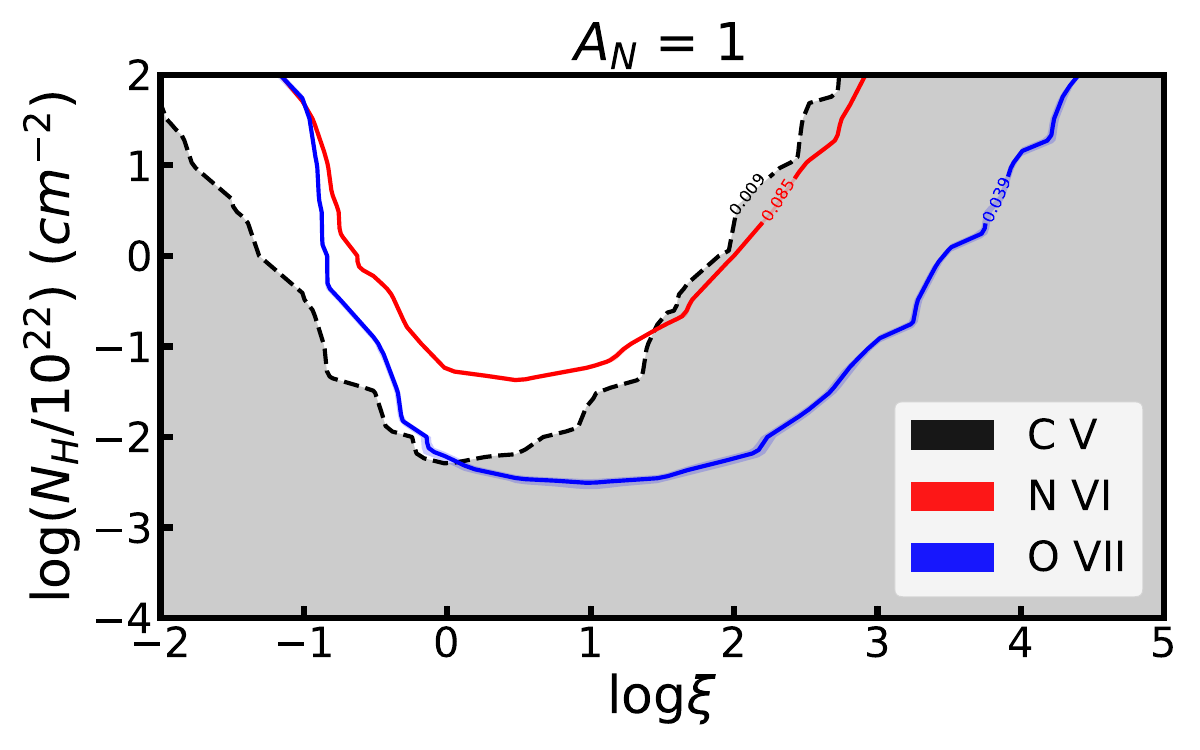}
    \caption{}
\end{subfigure}
\hfill
\begin{subfigure}[b]{0.45\textwidth}
    \centering
    \includegraphics[width=\textwidth]{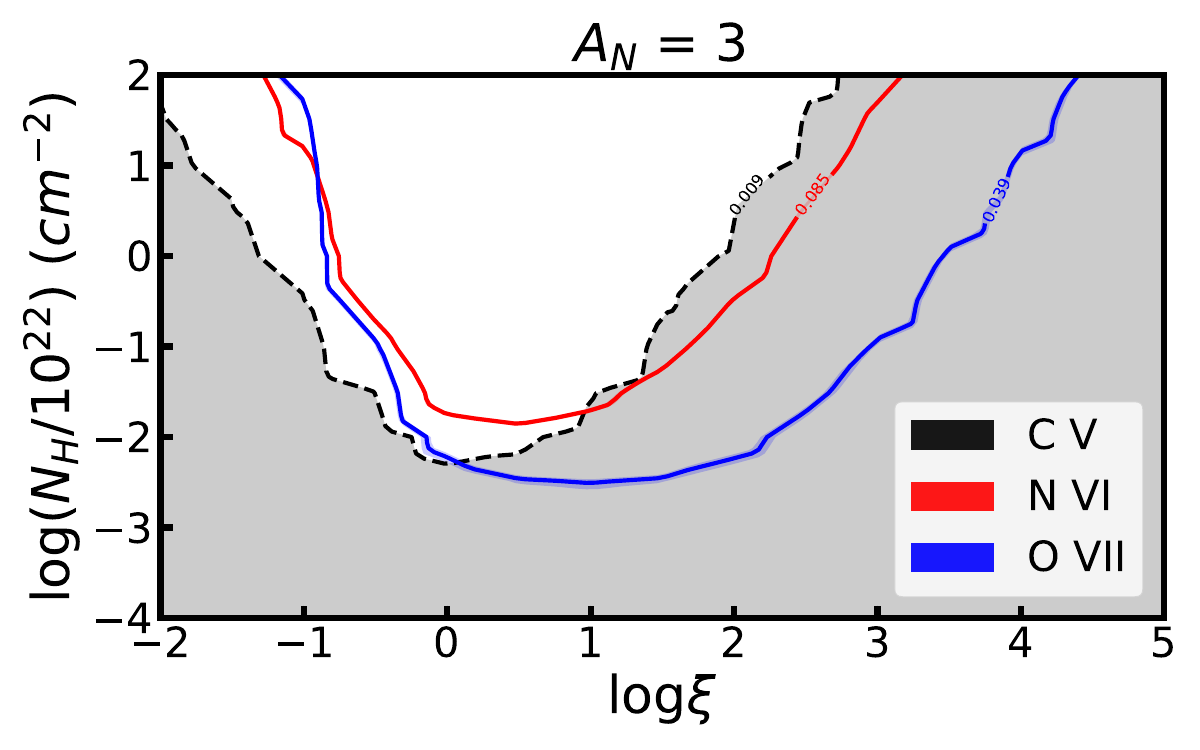}
    \caption{}
\end{subfigure}
\begin{subfigure}[b]{0.45\textwidth}
    \centering
    \includegraphics[width=\textwidth]{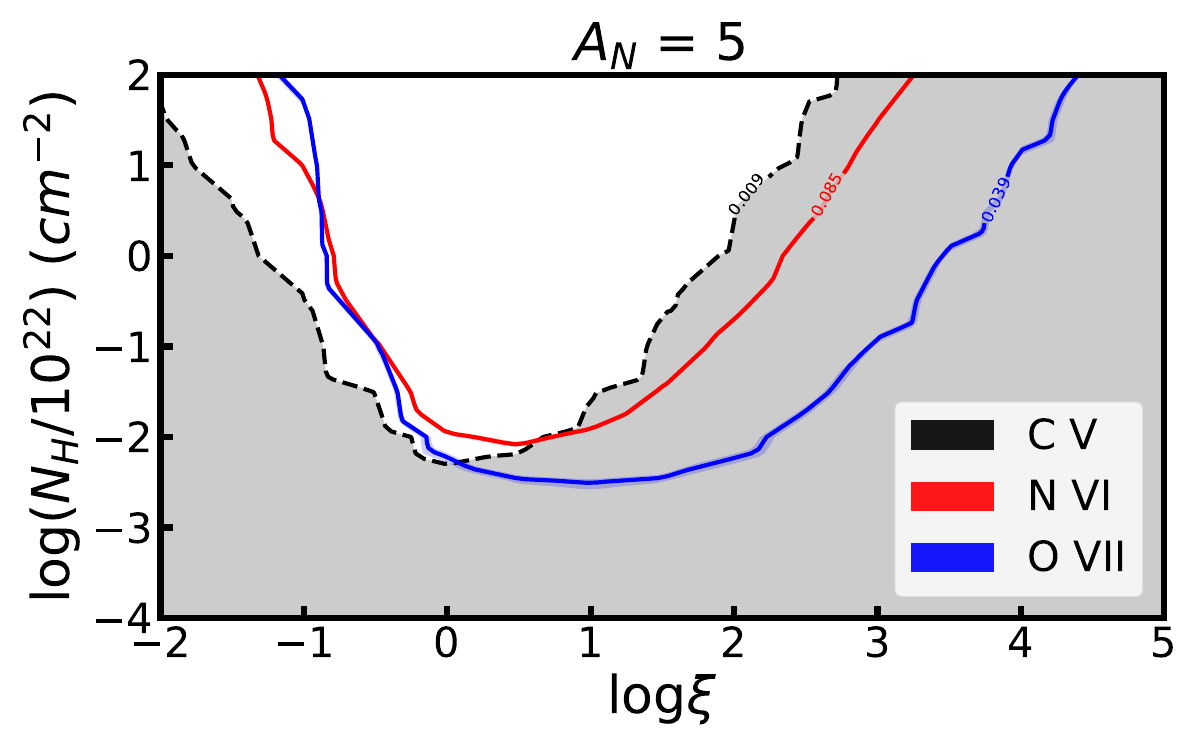}
    \caption{}
\end{subfigure}
\hfill
\begin{subfigure}[b]{0.45\textwidth}
    \centering
    \includegraphics[width=\textwidth]{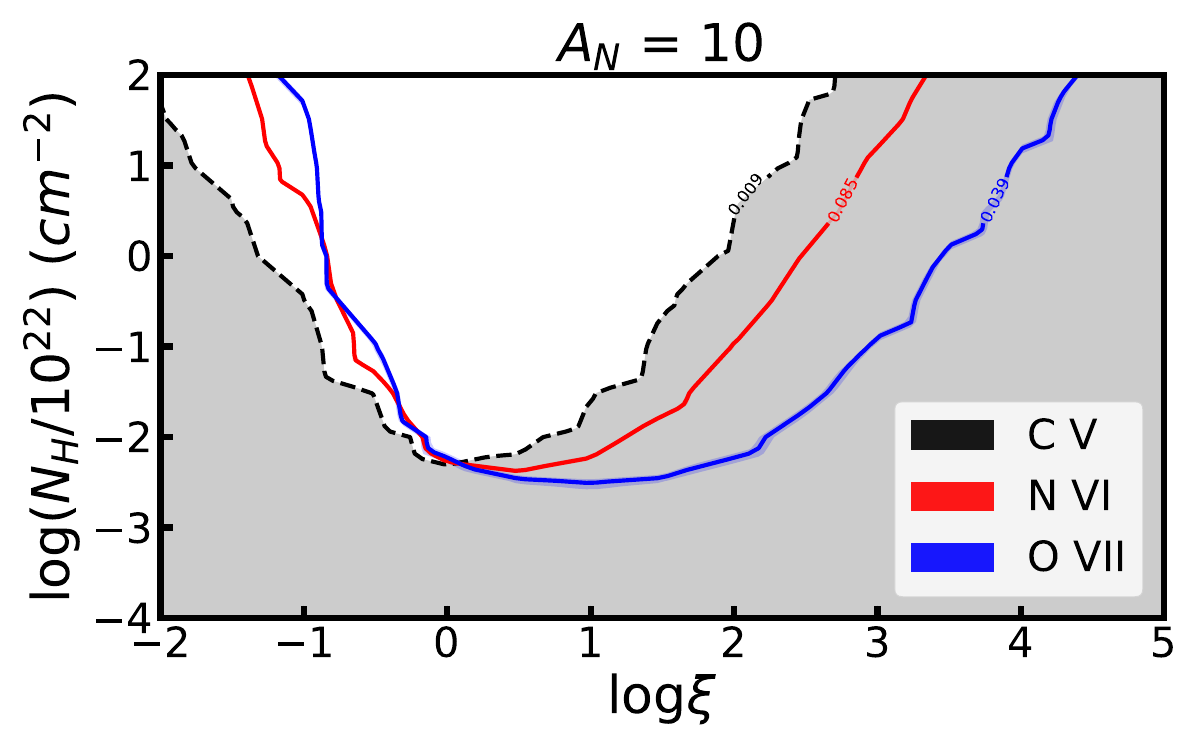}
    \caption{}
\end{subfigure}
\caption{
Panel (a): Simulated 100 ks HUBS spectrum of AT2019wzc, based on the best-fitting absorption model shown in Fig.~\ref{fig:pion}.
Panels (b)–(e): Corresponding constraints on the ionization parameter and column density of absorber,  
with nitrogen abundance increased 
from 1 to 10 times solar.
With a 100 ks HUBS observation, the equivalent widths of C {\sc v}, N {\sc vi}, and O {\sc vii} absorption lines and physical parameters of absorber are better constrained, and the nitrogen abundance can be reliably measured to be $A_{\rm N} \gtrsim 10$.
\label{fig:HUBS_pion_cal}
}
\end{figure*}

To quantify the improvements expected from HUBS,
we performed a spectral simulation with an exposure of 100 ks using its  latest available response files. 
In the simulation, we adopted the same best-fitting continuum and absorber model derived from our RGS analysis, including the enhanced N abundance.
Fig.~\ref{fig:HUBS_pion_cal}(a) shows the simulated HUBS spectrum. 
It can be seen that the N {\sc vi} absorption line is detected with a much higher signal-to-noise, and additional lines from O {\sc vii} and N {\sc vii} also become detectable.
These additional lines provide much stronger constraints on the N abundance and physical conditions of the absorber.
Following the method described in Section~\ref{subsec:physical conditions}, we measured the EWs of the C {\sc v}, N {\sc vi}, and O {\sc vii} absorption lines in the simulated spectrum, and set constraints on $\log \xi$, $N_H$ and $A_N$ using these EWs.
The results are shown in Fig.~\ref{fig:HUBS_pion_cal}(b)–(e), indicating that $A_N$ can be reliably determined to be $\gtrsim10$. 
This improved data quality will enable direct comparisons with stellar evolution models, thus allowing for further constraints on the disrupted star's properties \citep[e.g., ][]{Mockler2022}.
In addition, the kinematic parameters of the absorber are also much better constrained.
In comparison with the current {\it XMM-Newton}/RGS data, the uncertainties in the turbulent velocity and blueshifted velocity are reduced by factors of $\sim5$ and $\sim10$, respectively. 
Our simulation demonstrates that HUBS has the potential to uncover the anomaly of N abundance in AT2019wzc-like TDE candidates. 
This will help to confirm the TDE nature of ambiguous nuclear transients and shed new insights into the properties of disrupted stars.

\section{Conclusion} \label{sec:conclusion}

We analyzed the archival {\it XMM-Newton}/RGS high-resolution X-ray spectra of the quasi-periodically erupting source AT2019wzc, 
with a particular focus on its connection with a TDE. 
Our main results are summarized as follows:

\begin{enumerate} 

\item[$\bullet$] Through a blind line scan of the RGS spectrum, we find an isolated N {\sc vi} absorption line in the XMM2 data, with a statistical significance of $\sim 4\sigma$. 

\item[$\bullet$] The prominent N {\sc vi} absorption feature can be well described by an ionized absorber with enhanced nitrogen abundance. 
The absorber is characterized by a blueshifted and turbulent velocity of hundreds ${\rm km~s^{-1}}$. 
The best-fitting column density and ionization parameter are relatively low, with $N_{\rm H} = 0.0037_{-0.0032}^{+0.0054} \times 10^{22}~{\rm cm^{-2}}$ and $\log \xi = 0.3_{-0.3}^{+0.4}~{\rm erg~cm~s^{-1}}$. 
We further find that nitrogen is significantly over-abundant, with a best-fit value of $A_{\rm N} = 11.6_{-7.8}^{+19.6}$. 

\item[$\bullet$] Using extensive photoionization calculations and the observed equivalent widths of C {\sc v}, N {\sc vi}, and O {\sc vii} absorption lines, we find that to produce an isolated N {\sc vi} feature requires both enhanced nitrogen abundance ($A_N\gtrsim3-5$) and an extreme parameter space of $N_{\rm H}$ and $\log \xi$. 
This demonstrates that the isolated N {\sc vi} absorption line can serve not only as a good tracer of nitrogen enrichment and TDE process, but also as a unique probe of low-density, low-ionization gas environments. 

\item[$\bullet$] The enhanced nitrogen abundance detected in AT2019wzc provides a compelling evidence that the outburst in this system was triggered by a TDE rather than turn-on AGN. 
This strengthens the observational link between QPEs and TDEs, lending further support to the unified scenario \citep{Jiang2025}. 

\item[$\bullet$] Within the framework of TDE, 
the absorber might be related to the CIO produced by the self-collision of stellar debris streams. 
The origin of CIO in post-main-sequence star TDEs can self-consistently explain why the outflow gas in AT2019wzc have lower velocities and ionization parameters compared to other TDEs. 



\end{enumerate}

\begin{acknowledgements}
This research made use of the HEASARC online data archive services,
supported by NASA/GSFC, and is based on observations obtained
with XMM-Newton, an ESA science mission with instruments and
contributions directly funded by ESA Member States and NASA. 
We thank Prof. Junjie Mao for providing the response files of HUBS and useful discussions on the X-ray grating spectral analysis. 
The work is supported by 
the National Science Foundation of China (NSFC) through grant No. 12192220 and 12192221, and 
National Key R\&D Program of China (No. 2025YFF0511101). 
\end{acknowledgements}

%

\bibliographystyle{aa}
\bibliography{J1335}

@ARTICLE{Kara2018,
       author = {{Kara}, E. and {Dai}, L. and {Reynolds}, C.~S. and {Kallman}, T.},
        title = "{Ultrafast outflow in tidal disruption event ASASSN-14li}",
      journal = {\mnras},
         year = 2018,
        month = mar,
       volume = {474},
       number = {3},
        pages = {3593-3598},
          doi = {10.1093/mnras/stx3004},
archivePrefix = {arXiv},
       eprint = {1711.06090},
 primaryClass = {astro-ph.HE},
       adsurl = {https://ui.adsabs.harvard.edu/abs/2018MNRAS.474.3593K}
}

@ARTICLE{McKinney2015,
       author = {{McKinney}, Jonathan C. and {Dai}, Lixin and {Avara}, Mark J.},
        title = "{Efficiency of super-Eddington magnetically-arrested accretion}",
      journal = {\mnras},
         year = 2015,
        month = nov,
       volume = {454},
       number = {1},
        pages = {L6-L10},
          doi = {10.1093/mnrasl/slv115},
archivePrefix = {arXiv},
       eprint = {1508.02433},
 primaryClass = {astro-ph.HE},
       adsurl = {https://ui.adsabs.harvard.edu/abs/2015MNRAS.454L...6M}
}

@ARTICLE{Coughlin2014,
       author = {{Coughlin}, Eric R. and {Begelman}, Mitchell C.},
        title = "{Hyperaccretion during Tidal Disruption Events: Weakly Bound Debris Envelopes and Jets}",
      journal = {\apj},
         year = 2014,
        month = feb,
       volume = {781},
       number = {2},
          eid = {82},
        pages = {82},
          doi = {10.1088/0004-637X/781/2/82},
archivePrefix = {arXiv},
       eprint = {1312.5314},
 primaryClass = {astro-ph.HE},
       adsurl = {https://ui.adsabs.harvard.edu/abs/2014ApJ...781...82C}
}

@ARTICLE{HuangJiamu2023,
       author = {{Huang}, Jiamu and {Lin}, Douglas N.~C. and {Shields}, Gregory},
        title = "{Metal enrichment due to embedded stars in AGN discs}",
      journal = {\mnras},
         year = 2023,
        month = nov,
       volume = {525},
       number = {4},
        pages = {5702-5718},
          doi = {10.1093/mnras/stad2642},
archivePrefix = {arXiv},
       eprint = {2308.15761},
 primaryClass = {astro-ph.GA},
       adsurl = {https://ui.adsabs.harvard.edu/abs/2023MNRAS.525.5702H}
}

@ARTICLE{LiuXin2018,
       author = {{Liu}, Xin and {Dittmann}, Alexander and {Shen}, Yue and {Jiang}, Linhua},
        title = "{A Candidate Tidal Disruption Event in a Quasar at z = 2.359 from Abundance Ratio Variability}",
      journal = {\apj},
         year = 2018,
        month = may,
       volume = {859},
       number = {1},
          eid = {8},
        pages = {8},
          doi = {10.3847/1538-4357/aabb04},
archivePrefix = {arXiv},
       eprint = {1803.06362},
 primaryClass = {astro-ph.GA},
       adsurl = {https://ui.adsabs.harvard.edu/abs/2018ApJ...859....8L}
}

@ARTICLE{Isobe2025,
       author = {{Isobe}, Yuki and {Maiolino}, Roberto and {D'Eugenio}, Francesco and {Curti}, Mirko and {Ji}, Xihan and {Juod{\v{z}}balis}, Ignas and {Scholtz}, Jan and {Feltre}, Anne and {Charlot}, St{\'e}phane and {{\"U}bler}, Hannah and {J. Bunker}, Andrew and {Carniani}, Stefano and {Curtis-Lake}, Emma and {Ji}, Zhiyuan and {Kumari}, Nimisha and {Rinaldi}, Pierluigi and {Robertson}, Brant and {Willott}, Chris and {Witstok}, Joris},
        title = "{JADES: nitrogen enhancement in high-redshift broad-line active galactic nuclei}",
      journal = {\mnras},
         year = 2025,
        month = jul,
       volume = {541},
       number = {1},
        pages = {L71-L79},
          doi = {10.1093/mnrasl/slaf056},
archivePrefix = {arXiv},
       eprint = {2502.12091},
 primaryClass = {astro-ph.GA},
       adsurl = {https://ui.adsabs.harvard.edu/abs/2025MNRAS.541L..71I}
}

@ARTICLE{JiangLinhua2008,
       author = {{Jiang}, Linhua and {Fan}, Xiaohui and {Vestergaard}, M.},
        title = "{A Sample of Quasars with Strong Nitrogen Emission Lines from the Sloan Digital Sky Survey}",
      journal = {\apj},
         year = 2008,
        month = jun,
       volume = {679},
       number = {2},
        pages = {962-966},
          doi = {10.1086/587868},
archivePrefix = {arXiv},
       eprint = {0802.4234},
 primaryClass = {astro-ph},
       adsurl = {https://ui.adsabs.harvard.edu/abs/2008ApJ...679..962J}
}

@ARTICLE{Baldwin2003,
       author = {{Baldwin}, J.~A. and {Hamann}, F. and {Korista}, K.~T. and {Ferland}, G.~J. and {Dietrich}, M. and {Warner}, C.},
        title = "{Chemical Abundances in Broad Emission Line Regions: The ``Nitrogen-loud'' Quasi-Stellar Object Q0353-383}",
      journal = {\apj},
         year = 2003,
        month = feb,
       volume = {583},
       number = {2},
        pages = {649-658},
          doi = {10.1086/345449},
archivePrefix = {arXiv},
       eprint = {astro-ph/0210153},
 primaryClass = {astro-ph},
       adsurl = {https://ui.adsabs.harvard.edu/abs/2003ApJ...583..649B}
}

@ARTICLE{Kosec2023,
       author = {{Kosec}, P. and {Pasham}, D. and {Kara}, E. and {Tombesi}, F.},
        title = "{Discovery of a Variable Multiphase Outflow in the X-Ray-emitting Tidal Disruption Event ASASSN-20qc}",
      journal = {\apj},
         year = 2023,
        month = sep,
       volume = {954},
       number = {2},
          eid = {170},
        pages = {170},
          doi = {10.3847/1538-4357/aced87},
archivePrefix = {arXiv},
       eprint = {2308.05250},
 primaryClass = {astro-ph.HE},
       adsurl = {https://ui.adsabs.harvard.edu/abs/2023ApJ...954..170K}
}

@ARTICLE{Kosec2025,
       author = {{Kosec}, P. and {Kara}, E. and {Brenneman}, L. and {Chakraborty}, J. and {Giustini}, M. and {Miniutti}, G. and {Pinto}, C. and {Rogantini}, D. and {Arcodia}, R. and {Middleton}, M. and {Sacchi}, A.},
        title = "{Detection of a Highly Ionized Outflow in the Quasiperiodically Erupting Source GSN 069}",
      journal = {\apj},
         year = 2025,
        month = jan,
       volume = {978},
       number = {1},
          eid = {10},
        pages = {10},
          doi = {10.3847/1538-4357/ad9249},
archivePrefix = {arXiv},
       eprint = {2406.17105},
 primaryClass = {astro-ph.HE},
       adsurl = {https://ui.adsabs.harvard.edu/abs/2025ApJ...978...10K}
}

@ARTICLE{Lu_Bonnerot2020,
       author = {{Lu}, Wenbin and {Bonnerot}, Cl{\'e}ment},
        title = "{Self-intersection of the fallback stream in tidal disruption events}",
      journal = {\mnras},
         year = 2020,
        month = feb,
       volume = {492},
       number = {1},
        pages = {686-707},
          doi = {10.1093/mnras/stz3405},
archivePrefix = {arXiv},
       eprint = {1904.12018},
 primaryClass = {astro-ph.HE},
       adsurl = {https://ui.adsabs.harvard.edu/abs/2020MNRAS.492..686L}
}

@ARTICLE{Miller2015,
       author = {{Miller}, Jon M. and {Kaastra}, Jelle S. and {Miller}, M. Coleman and {Reynolds}, Mark T. and {Brown}, Gregory and {Cenko}, S. Bradley and {Drake}, Jeremy J. and {Gezari}, Suvi and {Guillochon}, James and {Gultekin}, Kayhan and {Irwin}, Jimmy and {Levan}, Andrew and {Maitra}, Dipankar and {Maksym}, W. Peter and {Mushotzky}, Richard and {O'Brien}, Paul and {Paerels}, Frits and {de Plaa}, Jelle and {Ramirez-Ruiz}, Enrico and {Strohmayer}, Tod and {Tanvir}, Nial},
        title = "{Flows of X-ray gas reveal the disruption of a star by a massive black hole}",
      journal = {\nat},
         year = 2015,
        month = oct,
       volume = {526},
       number = {7574},
        pages = {542-545},
          doi = {10.1038/nature15708},
archivePrefix = {arXiv},
       eprint = {1510.06348},
 primaryClass = {astro-ph.HE},
       adsurl = {https://ui.adsabs.harvard.edu/abs/2015Natur.526..542M}
}

@ARTICLE{Tombesi2013,
       author = {{Tombesi}, F. and {Cappi}, M. and {Reeves}, J.~N. and {Nemmen}, R.~S. and {Braito}, V. and {Gaspari}, M. and {Reynolds}, C.~S.},
        title = "{Unification of X-ray winds in Seyfert galaxies: from ultra-fast outflows to warm absorbers}",
      journal = {\mnras},
         year = 2013,
        month = apr,
       volume = {430},
       number = {2},
        pages = {1102-1117},
          doi = {10.1093/mnras/sts692},
archivePrefix = {arXiv},
       eprint = {1212.4851},
 primaryClass = {astro-ph.HE},
       adsurl = {https://ui.adsabs.harvard.edu/abs/2013MNRAS.430.1102T}
}

@ARTICLE{Kejriwal2024,
       author = {{Kejriwal}, Shubham and {Witzany}, Vojt{\v{e}}ch and {Zaja{\v{c}}ek}, Michal and {Pasham}, Dheeraj R. and {Chua}, Alvin J.~K.},
        title = "{Repeating nuclear transients as candidate electromagnetic counterparts of LISA extreme mass ratio inspirals}",
      journal = {\mnras},
         year = 2024,
        month = aug,
       volume = {532},
       number = {2},
        pages = {2143-2158},
          doi = {10.1093/mnras/stae1599},
archivePrefix = {arXiv},
       eprint = {2404.00941},
 primaryClass = {astro-ph.HE},
       adsurl = {https://ui.adsabs.harvard.edu/abs/2024MNRAS.532.2143K}
}

@ARTICLE{King2022,
       author = {{King}, Andrew},
        title = "{Quasi-periodic eruptions from galaxy nuclei}",
      journal = {\mnras},
         year = 2022,
        month = sep,
       volume = {515},
       number = {3},
        pages = {4344-4349},
          doi = {10.1093/mnras/stac1641},
archivePrefix = {arXiv},
       eprint = {2206.04698},
 primaryClass = {astro-ph.GA},
       adsurl = {https://ui.adsabs.harvard.edu/abs/2022MNRAS.515.4344K}
}

@ARTICLE{Sukova2021,
       author = {{Sukov{\'a}}, Petra and {Zaja{\v{c}}ek}, Michal and {Witzany}, Vojt{\v{e}}ch and {Karas}, Vladim{\'\i}r},
        title = "{Stellar Transits across a Magnetized Accretion Torus as a Mechanism for Plasmoid Ejection}",
      journal = {\apj},
         year = 2021,
        month = aug,
       volume = {917},
       number = {1},
          eid = {43},
        pages = {43},
          doi = {10.3847/1538-4357/ac05c6},
archivePrefix = {arXiv},
       eprint = {2102.08135},
 primaryClass = {astro-ph.HE},
       adsurl = {https://ui.adsabs.harvard.edu/abs/2021ApJ...917...43S}
}

@ARTICLE{Dai2010,
       author = {{Dai}, Lixin Jane and {Fuerst}, Steven V. and {Blandford}, Roger},
        title = "{Quasi-periodic flares from star-accretion-disc collisions}",
      journal = {\mnras},
         year = 2010,
        month = mar,
       volume = {402},
       number = {3},
        pages = {1614-1624},
          doi = {10.1111/j.1365-2966.2009.16038.x},
archivePrefix = {arXiv},
       eprint = {0906.0800},
 primaryClass = {astro-ph.HE},
       adsurl = {https://ui.adsabs.harvard.edu/abs/2010MNRAS.402.1614D}
}

@ARTICLE{Bykov2025,
       author = {{Bykov}, S.~D. and {Gilfanov}, M.~R. and {Sunyaev}, R.~A. and {Medvedev}, P.~S.},
        title = "{Further evidence of quasi-periodic eruptions in a tidal disruption event AT2019vcb by SRG/eROSITA}",
      journal = {\mnras},
         year = 2025,
        month = jun,
       volume = {540},
       number = {1},
        pages = {30-36},
          doi = {10.1093/mnras/staf686},
archivePrefix = {arXiv},
       eprint = {2409.16908},
 primaryClass = {astro-ph.HE},
       adsurl = {https://ui.adsabs.harvard.edu/abs/2025MNRAS.540...30B}
}

@ARTICLE{Mockler2022,
       author = {{Mockler}, Brenna and {Twum}, Angela A. and {Auchettl}, Katie and {Dodd}, Sierra and {French}, K.~D. and {Law-Smith}, Jamie A.~P. and {Ramirez-Ruiz}, Enrico},
        title = "{Evidence for the Preferential Disruption of Moderately Massive Stars by Supermassive Black Holes}",
      journal = {\apj},
         year = 2022,
        month = jan,
       volume = {924},
       number = {2},
          eid = {70},
        pages = {70},
          doi = {10.3847/1538-4357/ac35d5},
archivePrefix = {arXiv},
       eprint = {2110.03013},
 primaryClass = {astro-ph.HE},
       adsurl = {https://ui.adsabs.harvard.edu/abs/2022ApJ...924...70M}
}

@ARTICLE{Lambert1981,
       author = {{Lambert}, D.~L. and {Ries}, L.~M.},
        title = "{Carbon, nitrogen, and oxygen abundances in G and K giants.}",
      journal = {\apj},
         year = 1981,
        month = aug,
       volume = {248},
        pages = {228-248},
          doi = {10.1086/159147},
       adsurl = {https://ui.adsabs.harvard.edu/abs/1981ApJ...248..228L}
}

@ARTICLE{Lambert1977,
       author = {{Lambert}, D.~L. and {Ries}, L.~M.},
        title = "{Carbon, nitrogen, and oxygen abundances in 11 G and K giants.}",
      journal = {\apj},
         year = 1977,
        month = oct,
       volume = {217},
        pages = {508-520},
          doi = {10.1086/155599},
       adsurl = {https://ui.adsabs.harvard.edu/abs/1977ApJ...217..508L}
}

@ARTICLE{Iben1967,
       author = {{Iben}, Jr., Icko},
        title = "{Stellar Evolution.VI. Evolution from the Main Sequence to the Red-Giant Branch for Stars of Mass 1 M\_\{sun\}, 1.25 M\_\{sun\}, and 1.5 M\_\{sun\}}",
      journal = {\apj},
         year = 1967,
        month = feb,
       volume = {147},
        pages = {624},
          doi = {10.1086/149040},
       adsurl = {https://ui.adsabs.harvard.edu/abs/1967ApJ...147..624I}
}

@ARTICLE{Iben1964,
       author = {{Iben}, Jr., Icko},
        title = "{The Surface Ratio of N$^{14}$ to C$^{12}$ during Helium Burning.}",
      journal = {\apj},
         year = 1964,
        month = nov,
       volume = {140},
        pages = {1631},
          doi = {10.1086/148077},
       adsurl = {https://ui.adsabs.harvard.edu/abs/1964ApJ...140.1631I}
}

@ARTICLE{Mockler2024,
       author = {{Mockler}, Brenna and {Gallegos-Garcia}, Monica and {G{\"o}tberg}, Ylva and {Miller}, Jon M. and {Ramirez-Ruiz}, Enrico},
        title = "{Tidal Disruption Events from Stripped Stars}",
      journal = {\apjl},
         year = 2024,
        month = sep,
       volume = {973},
       number = {1},
          eid = {L9},
        pages = {L9},
          doi = {10.3847/2041-8213/ad6c34},
archivePrefix = {arXiv},
       eprint = {2406.04455},
 primaryClass = {astro-ph.HE},
       adsurl = {https://ui.adsabs.harvard.edu/abs/2024ApJ...973L...9M}
}

@ARTICLE{Kochanek2016,
       author = {{Kochanek}, C.~S.},
        title = "{Abundance anomalies in tidal disruption events}",
      journal = {\mnras},
         year = 2016,
        month = may,
       volume = {458},
       number = {1},
        pages = {127-134},
          doi = {10.1093/mnras/stw267},
archivePrefix = {arXiv},
       eprint = {1512.03065},
 primaryClass = {astro-ph.HE},
       adsurl = {https://ui.adsabs.harvard.edu/abs/2016MNRAS.458..127K}
}

@ARTICLE{Cenko2016,
       author = {{Cenko}, S. Bradley and {Cucchiara}, Antonino and {Roth}, Nathaniel and {Veilleux}, Sylvain and {Prochaska}, J. Xavier and {Yan}, Lin and {Guillochon}, James and {Maksym}, W. Peter and {Arcavi}, Iair and {Butler}, Nathaniel R. and {Filippenko}, Alexei V. and {Fruchter}, Andrew S. and {Gezari}, Suvi and {Kasen}, Daniel and {Levan}, Andrew J. and {Miller}, Jon M. and {Pasham}, Dheeraj R. and {Ramirez-Ruiz}, Enrico and {Strubbe}, Linda E. and {Tanvir}, Nial R. and {Tombesi}, Francesco},
        title = "{An Ultraviolet Spectrum of the Tidal Disruption Flare ASASSN-14li}",
      journal = {\apjl},
         year = 2016,
        month = feb,
       volume = {818},
       number = {2},
          eid = {L32},
        pages = {L32},
          doi = {10.3847/2041-8205/818/2/L32},
archivePrefix = {arXiv},
       eprint = {1601.03331},
 primaryClass = {astro-ph.HE},
       adsurl = {https://ui.adsabs.harvard.edu/abs/2016ApJ...818L..32C}
}

@ARTICLE{denHerder2001,
       author = {{den Herder}, J.~W. and {Brinkman}, A.~C. and {Kahn}, S.~M. and {Branduardi-Raymont}, G. and {Thomsen}, K. and {Aarts}, H. and {Audard}, M. and {Bixler}, J.~V. and {den Boggende}, A.~J. and {Cottam}, J. and {Decker}, T. and {Dubbeldam}, L. and {Erd}, C. and {Goulooze}, H. and {G{\"u}del}, M. and {Guttridge}, P. and {Hailey}, C.~J. and {Janabi}, K. Al and {Kaastra}, J.~S. and {de Korte}, P.~A.~J. and {van Leeuwen}, B.~J. and {Mauche}, C. and {McCalden}, A.~J. and {Mewe}, R. and {Naber}, A. and {Paerels}, F.~B. and {Peterson}, J.~R. and {Rasmussen}, A.~P. and {Rees}, K. and {Sakelliou}, I. and {Sako}, M. and {Spodek}, J. and {Stern}, M. and {Tamura}, T. and {Tandy}, J. and {de Vries}, C.~P. and {Welch}, S. and {Zehnder}, A.},
        title = "{The Reflection Grating Spectrometer on board XMM-Newton}",
      journal = {\aap},
         year = 2001,
        month = jan,
       volume = {365},
        pages = {L7-L17},
          doi = {10.1051/0004-6361:20000058},
       adsurl = {https://ui.adsabs.harvard.edu/abs/2001A&A...365L...7D}
}

@ARTICLE{Jansen2001,
       author = {{Jansen}, F. and {Lumb}, D. and {Altieri}, B. and {Clavel}, J. and {Ehle}, M. and {Erd}, C. and {Gabriel}, C. and {Guainazzi}, M. and {Gondoin}, P. and {Much}, R. and {Munoz}, R. and {Santos}, M. and {Schartel}, N. and {Texier}, D. and {Vacanti}, G.},
        title = "{XMM-Newton observatory. I. The spacecraft and operations}",
      journal = {\aap},
         year = 2001,
        month = jan,
       volume = {365},
        pages = {L1-L6},
          doi = {10.1051/0004-6361:20000036},
       adsurl = {https://ui.adsabs.harvard.edu/abs/2001A&A...365L...1J}
}

@ARTICLE{Yang2017,
       author = {{Yang}, Chenwei and {Wang}, Tinggui and {Ferland}, Gary J. and {Dou}, Liming and {Zhou}, Hongyan and {Jiang}, Ning and {Sheng}, Zhenfeng},
        title = "{The Carbon and Nitrogen Abundance Ratio in the Broad Line Region of Tidal Disruption Events}",
      journal = {\apj},
         year = 2017,
        month = sep,
       volume = {846},
       number = {2},
          eid = {150},
        pages = {150},
          doi = {10.3847/1538-4357/aa8598},
archivePrefix = {arXiv},
       eprint = {1708.03548},
 primaryClass = {astro-ph.GA},
       adsurl = {https://ui.adsabs.harvard.edu/abs/2017ApJ...846..150Y}
}

@ARTICLE{Rees1988,
       author = {{Rees}, Martin J.},
        title = "{Tidal disruption of stars by black holes of {}10$^{6}$-{}10$^{8}$ solar masses in nearby galaxies}",
      journal = {\nat},
         year = 1988,
        month = jun,
       volume = {333},
       number = {6173},
        pages = {523-528},
          doi = {10.1038/333523a0},
       adsurl = {https://ui.adsabs.harvard.edu/abs/1988Natur.333..523R}
}

@ARTICLE{Wang2022,
       author = {{Wang}, Mengye and {Yin}, Jinjing and {Ma}, Yiqiu and {Wu}, Qingwen},
        title = "{A Model for the Possible Connection Between a Tidal Disruption Event and Quasi-periodic Eruption in GSN 069}",
      journal = {\apj},
         year = 2022,
        month = jul,
       volume = {933},
       number = {2},
          eid = {225},
        pages = {225},
          doi = {10.3847/1538-4357/ac75e6},
archivePrefix = {arXiv},
       eprint = {2206.03092},
 primaryClass = {astro-ph.HE},
       adsurl = {https://ui.adsabs.harvard.edu/abs/2022ApJ...933..225W}
}

@ARTICLE{Zhao2022,
       author = {{Zhao}, Z.~Y. and {Wang}, Y.~Y. and {Zou}, Y.~C. and {Wang}, F.~Y. and {Dai}, Z.~G.},
        title = "{Quasi-periodic eruptions from the helium envelope of hydrogen-deficient stars stripped by supermassive black holes}",
      journal = {\aap},
         year = 2022,
        month = may,
       volume = {661},
          eid = {A55},
        pages = {A55},
          doi = {10.1051/0004-6361/202142519},
archivePrefix = {arXiv},
       eprint = {2109.03471},
 primaryClass = {astro-ph.HE},
       adsurl = {https://ui.adsabs.harvard.edu/abs/2022A&A...661A..55Z}
}

@ARTICLE{Hammerstein2023,
       author = {{Hammerstein}, Erica and {van Velzen}, Sjoert and {Gezari}, Suvi and {Cenko}, S. Bradley and {Yao}, Yuhan and {Ward}, Charlotte and {Frederick}, Sara and {Villanueva}, Natalia and {Somalwar}, Jean J. and {Graham}, Matthew J. and {Kulkarni}, Shrinivas R. and {Stern}, Daniel and {Andreoni}, Igor and {Bellm}, Eric C. and {Dekany}, Richard and {Dhawan}, Suhail and {Drake}, Andrew J. and {Fremling}, Christoffer and {Gatkine}, Pradip and {Groom}, Steven L. and {Ho}, Anna Y.~Q. and {Kasliwal}, Mansi M. and {Karambelkar}, Viraj and {Kool}, Erik C. and {Masci}, Frank J. and {Medford}, Michael S. and {Perley}, Daniel A. and {Purdum}, Josiah and {van Roestel}, Jan and {Sharma}, Yashvi and {Sollerman}, Jesper and {Taggart}, Kirsty and {Yan}, Lin},
        title = "{The Final Season Reimagined: 30 Tidal Disruption Events from the ZTF-I Survey}",
      journal = {\apj},
         year = 2023,
        month = jan,
       volume = {942},
       number = {1},
          eid = {9},
        pages = {9},
          doi = {10.3847/1538-4357/aca283},
archivePrefix = {arXiv},
       eprint = {2203.01461},
 primaryClass = {astro-ph.HE},
       adsurl = {https://ui.adsabs.harvard.edu/abs/2023ApJ...942....9H}
}

@ARTICLE{Shu2018,
       author = {{Shu}, X.~W. and {Wang}, S.~S. and {Dou}, L.~M. and {Jiang}, N. and {Wang}, J.~X. and {Wang}, T.~G.},
        title = "{A Long Decay of X-Ray Flux and Spectral Evolution in the Supersoft Active Galactic Nucleus GSN 069}",
      journal = {\apjl},
         year = 2018,
        month = apr,
       volume = {857},
       number = {2},
          eid = {L16},
        pages = {L16},
          doi = {10.3847/2041-8213/aaba17},
archivePrefix = {arXiv},
       eprint = {1809.00319},
 primaryClass = {astro-ph.HE},
       adsurl = {https://ui.adsabs.harvard.edu/abs/2018ApJ...857L..16S}
}

@ARTICLE{Miniutti2019,
       author = {{Miniutti}, G. and {Saxton}, R.~D. and {Giustini}, M. and {Alexander}, K.~D. and {Fender}, R.~P. and {Heywood}, I. and {Monageng}, I. and {Coriat}, M. and {Tzioumis}, A.~K. and {Read}, A.~M. and {Knigge}, C. and {Gandhi}, P. and {Pretorius}, M.~L. and {Ag{\'\i}s-Gonz{\'a}lez}, B.},
        title = "{Nine-hour X-ray quasi-periodic eruptions from a low-mass black hole galactic nucleus}",
      journal = {\nat},
         year = 2019,
        month = sep,
       volume = {573},
       number = {7774},
        pages = {381-384},
          doi = {10.1038/s41586-019-1556-x},
archivePrefix = {arXiv},
       eprint = {1909.04693},
 primaryClass = {astro-ph.HE},
       adsurl = {https://ui.adsabs.harvard.edu/abs/2019Natur.573..381M}
}

@ARTICLE{Sun2013,
       author = {{Sun}, Luming and {Shu}, Xinwen and {Wang}, Tinggui},
        title = "{RX J1301.9+2747: A Highly Variable Seyfert Galaxy with Extremely Soft X-Ray Emission}",
      journal = {\apj},
         year = 2013,
        month = may,
       volume = {768},
       number = {2},
          eid = {167},
        pages = {167},
          doi = {10.1088/0004-637X/768/2/167},
archivePrefix = {arXiv},
       eprint = {1304.3244},
 primaryClass = {astro-ph.GA},
       adsurl = {https://ui.adsabs.harvard.edu/abs/2013ApJ...768..167S}
}

@ARTICLE{Giustini2020,
       author = {{Giustini}, Margherita and {Miniutti}, Giovanni and {Saxton}, Richard D.},
        title = "{X-ray quasi-periodic eruptions from the galactic nucleus of RX J1301.9+2747}",
      journal = {\aap},
         year = 2020,
        month = apr,
       volume = {636},
          eid = {L2},
        pages = {L2},
          doi = {10.1051/0004-6361/202037610},
archivePrefix = {arXiv},
       eprint = {2002.08967},
 primaryClass = {astro-ph.HE},
       adsurl = {https://ui.adsabs.harvard.edu/abs/2020A&A...636L...2G}
}

@ARTICLE{Arcodia2021,
       author = {{Arcodia}, R. and {Merloni}, A. and {Nandra}, K. and {Buchner}, J. and {Salvato}, M. and {Pasham}, D. and {Remillard}, R. and {Comparat}, J. and {Lamer}, G. and {Ponti}, G. and {Malyali}, A. and {Wolf}, J. and {Arzoumanian}, Z. and {Bogensberger}, D. and {Buckley}, D.~A.~H. and {Gendreau}, K. and {Gromadzki}, M. and {Kara}, E. and {Krumpe}, M. and {Markwardt}, C. and {Ramos-Ceja}, M.~E. and {Rau}, A. and {Schramm}, M. and {Schwope}, A.},
        title = "{X-ray quasi-periodic eruptions from two previously quiescent galaxies}",
      journal = {\nat},
         year = 2021,
        month = apr,
       volume = {592},
       number = {7856},
        pages = {704-707},
          doi = {10.1038/s41586-021-03394-6},
archivePrefix = {arXiv},
       eprint = {2104.13388},
 primaryClass = {astro-ph.HE},
       adsurl = {https://ui.adsabs.harvard.edu/abs/2021Natur.592..704A}
}

@ARTICLE{Chakraborty2021,
       author = {{Chakraborty}, Joheen and {Kara}, Erin and {Masterson}, Megan and {Giustini}, Margherita and {Miniutti}, Giovanni and {Saxton}, Richard},
        title = "{Possible X-Ray Quasi-periodic Eruptions in a Tidal Disruption Event Candidate}",
      journal = {\apjl},
         year = 2021,
        month = nov,
       volume = {921},
       number = {2},
          eid = {L40},
        pages = {L40},
          doi = {10.3847/2041-8213/ac313b},
archivePrefix = {arXiv},
       eprint = {2110.10786},
 primaryClass = {astro-ph.HE},
       adsurl = {https://ui.adsabs.harvard.edu/abs/2021ApJ...921L..40C}
}

@ARTICLE{Quintin2023,
       author = {{Quintin}, E. and {Webb}, N.~A. and {Guillot}, S. and {Miniutti}, G. and {Kammoun}, E.~S. and {Giustini}, M. and {Arcodia}, R. and {Soucail}, G. and {Clerc}, N. and {Amato}, R. and {Markwardt}, C.~B.},
        title = "{Tormund's return: Hints of quasi-periodic eruption features from a recent optical tidal disruption event}",
      journal = {\aap},
         year = 2023,
        month = jul,
       volume = {675},
          eid = {A152},
        pages = {A152},
          doi = {10.1051/0004-6361/202346440},
archivePrefix = {arXiv},
       eprint = {2306.00438},
 primaryClass = {astro-ph.HE},
       adsurl = {https://ui.adsabs.harvard.edu/abs/2023A&A...675A.152Q}
}

@ARTICLE{Arcodia2024,
       author = {{Arcodia}, R. and {Liu}, Z. and {Merloni}, A. and {Malyali}, A. and {Rau}, A. and {Chakraborty}, J. and {Goodwin}, A. and {Buckley}, D. and {Brink}, J. and {Gromadzki}, M. and {Arzoumanian}, Z. and {Buchner}, J. and {Kara}, E. and {Nandra}, K. and {Ponti}, G. and {Salvato}, M. and {Anderson}, G. and {Baldini}, P. and {Grotova}, I. and {Krumpe}, M. and {Maitra}, C. and {Miller-Jones}, J.~C.~A. and {Ramos-Ceja}, M.~E.},
        title = "{The more the merrier: SRG/eROSITA discovers two further galaxies showing X-ray quasi-periodic eruptions}",
      journal = {\aap},
         year = 2024,
        month = apr,
       volume = {684},
          eid = {A64},
        pages = {A64},
          doi = {10.1051/0004-6361/202348881},
archivePrefix = {arXiv},
       eprint = {2401.17275},
 primaryClass = {astro-ph.HE},
       adsurl = {https://ui.adsabs.harvard.edu/abs/2024A&A...684A..64A}
}

@ARTICLE{Nicholl2024,
       author = {{Nicholl}, M. and {Pasham}, D.~R. and {Mummery}, A. and {Guolo}, M. and {Gendreau}, K. and {Dewangan}, G.~C. and {Ferrara}, E.~C. and {Remillard}, R. and {Bonnerot}, C. and {Chakraborty}, J. and {Hajela}, A. and {Dhillon}, V.~S. and {Gillan}, A.~F. and {Greenwood}, J. and {Huber}, M.~E. and {Janiuk}, A. and {Salvesen}, G. and {van Velzen}, S. and {Aamer}, A. and {Alexander}, K.~D. and {Angus}, C.~R. and {Arzoumanian}, Z. and {Auchettl}, K. and {Berger}, E. and {de Boer}, T. and {Cendes}, Y. and {Chambers}, K.~C. and {Chen}, T. -W. and {Chornock}, R. and {Fulton}, M.~D. and {Gao}, H. and {Gillanders}, J.~H. and {Gomez}, S. and {Gompertz}, B.~P. and {Fabian}, A.~C. and {Herman}, J. and {Ingram}, A. and {Kara}, E. and {Laskar}, T. and {Lawrence}, A. and {Lin}, C. -C. and {Lowe}, T.~B. and {Magnier}, E.~A. and {Margutti}, R. and {McGee}, S.~L. and {Minguez}, P. and {Moore}, T. and {Nathan}, E. and {Oates}, S.~R. and {Patra}, K.~C. and {Ramsden}, P. and {Ravi}, V. and {Ridley}, E.~J. and {Sheng}, X. and {Smartt}, S.~J. and {Smith}, K.~W. and {Srivastav}, S. and {Stein}, R. and {Stevance}, H.~F. and {Turner}, S.~G.~D. and {Wainscoat}, R.~J. and {Weston}, J. and {Wevers}, T. and {Young}, D.~R.},
        title = "{Quasi-periodic X-ray eruptions years after a nearby tidal disruption event}",
      journal = {\nat},
         year = 2024,
        month = oct,
       volume = {634},
       number = {8035},
        pages = {804-808},
          doi = {10.1038/s41586-024-08023-6},
archivePrefix = {arXiv},
       eprint = {2409.02181},
 primaryClass = {astro-ph.HE},
       adsurl = {https://ui.adsabs.harvard.edu/abs/2024Natur.634..804N}
}

@ARTICLE{Chakraborty2025b,
       author = {{Chakraborty}, Joheen and {Kosec}, Peter and {Kara}, Erin and {Miniutti}, Giovanni and {Arcodia}, Riccardo and {Behar}, Ehud and {Giustini}, Margherita and {Hern{\'a}ndez-Garc{\'\i}a}, Lorena and {Masterson}, Megan and {Quintin}, Erwan and {Ricci}, Claudio and {S{\'a}nchez-S{\'a}ez}, Paula},
        title = "{Rapidly Varying Ionization Features in a Quasi-periodic Eruption: A Homologous Expansion Model for the Spectroscopic Evolution}",
      journal = {\apj},
         year = 2025,
        month = may,
       volume = {984},
       number = {2},
          eid = {124},
        pages = {124},
          doi = {10.3847/1538-4357/adb972},
archivePrefix = {arXiv},
       eprint = {2504.07167},
 primaryClass = {astro-ph.HE},
       adsurl = {https://ui.adsabs.harvard.edu/abs/2025ApJ...984..124C}
}

@ARTICLE{Chakraborty2025a,
       author = {{Chakraborty}, Joheen and {Kara}, Erin and {Arcodia}, Riccardo and {Buchner}, Johannes and {Giustini}, Margherita and {Hern{\'a}ndez-Garc{\'\i}a}, Lorena and {Linial}, Itai and {Masterson}, Megan and {Miniutti}, Giovanni and {Mummery}, Andrew and {Panagiotou}, Christos and {Quintin}, Erwan and {S{\'a}nchez-S{\'a}ez}, Paula},
        title = "{Discovery of Quasiperiodic Eruptions in the Tidal Disruption Event and Extreme Coronal Line Emitter AT2022upj: Implications for the QPE/TDE Fraction and a Connection to ECLEs}",
      journal = {\apjl},
         year = 2025,
        month = apr,
       volume = {983},
       number = {2},
          eid = {L39},
        pages = {L39},
          doi = {10.3847/2041-8213/adc2f8},
archivePrefix = {arXiv},
       eprint = {2503.19013},
 primaryClass = {astro-ph.HE},
       adsurl = {https://ui.adsabs.harvard.edu/abs/2025ApJ...983L..39C}
}

@ARTICLE{Arcodia2025,
       author = {{Arcodia}, R. and {Baldini}, P. and {Merloni}, A. and {Rau}, A. and {Nandra}, K. and {Chakraborty}, J. and {Goodwin}, A.~J. and {Page}, M.~J. and {Buchner}, J. and {Masterson}, M. and {Monageng}, I. and {Arzoumanian}, Z. and {Buckley}, D. and {Kara}, E. and {Ponti}, G. and {Ramos-Ceja}, M.~E. and {Salvato}, M. and {Gendreau}, K. and {Grotova}, I. and {Krumpe}, M.},
        title = "{SRG/eROSITA No. 5: Discovery of Quasiperiodic Eruptions Every {\ensuremath{\sim}}3.7 days from a Galaxy at z > 0.1}",
      journal = {\apj},
         year = 2025,
        month = aug,
       volume = {989},
       number = {1},
          eid = {13},
        pages = {13},
          doi = {10.3847/1538-4357/adec9b},
       adsurl = {https://ui.adsabs.harvard.edu/abs/2025ApJ...989...13A}
}

@ARTICLE{Hernandez-Garcia2025,
       author = {{Hern{\'a}ndez-Garc{\'\i}a}, Lorena and {Chakraborty}, Joheen and {S{\'a}nchez-S{\'a}ez}, Paula and {Ricci}, Claudio and {Cuadra}, Jorge and {McKernan}, Barry and {Ford}, K.~E. Saavik and {Ar{\'e}valo}, Patricia and {Rau}, Arne and {Arcodia}, Riccardo and {Kara}, Erin and {Liu}, Zhu and {Merloni}, Andrea and {Bruni}, Gabriele and {Goodwin}, Adelle and {Arzoumanian}, Zaven and {Assef}, Roberto J. and {Baldini}, Pietro and {Bayo}, Amelia and {Bauer}, Franz E. and {Bernal}, Santiago and {Brightman}, Murray and {Calistro Rivera}, Gabriela and {Gendreau}, Keith and {Homan}, David and {Krumpe}, Mirko and {Lira}, Paulina and {Mart{\'\i}nez-Aldama}, Mary Loli and {Salvato}, Mara and {Sotomayor}, Bel{\'e}n},
        title = "{Discovery of extreme quasi-periodic eruptions in a newly accreting massive black hole}",
      journal = {Nature Astronomy},
         year = 2025,
        month = apr,
          doi = {10.1038/s41550-025-02523-9},
       adsurl = {https://ui.adsabs.harvard.edu/abs/2025NatAs.tmp...90H}
}

@ARTICLE{Sheng2021,
       author = {{Sheng}, Zhenfeng and {Wang}, Tinggui and {Ferland}, Gary and {Shu}, Xinwen and {Yang}, Chenwei and {Jiang}, Ning and {Chen}, Yang},
        title = "{Evidence of a Tidal-disruption Event in GSN 069 from the Abnormal Carbon and Nitrogen Abundance Ratio}",
      journal = {\apjl},
         year = 2021,
        month = oct,
       volume = {920},
       number = {1},
          eid = {L25},
        pages = {L25},
          doi = {10.3847/2041-8213/ac2251},
archivePrefix = {arXiv},
       eprint = {2109.01683},
 primaryClass = {astro-ph.GA},
       adsurl = {https://ui.adsabs.harvard.edu/abs/2021ApJ...920L..25S}
}

@ARTICLE{king2020,
       author = {{King}, Andrew},
        title = "{GSN 069 - A tidal disruption near miss}",
      journal = {\mnras},
         year = 2020,
        month = mar,
       volume = {493},
       number = {1},
        pages = {L120-L123},
          doi = {10.1093/mnrasl/slaa020},
archivePrefix = {arXiv},
       eprint = {2002.00970},
 primaryClass = {astro-ph.HE},
       adsurl = {https://ui.adsabs.harvard.edu/abs/2020MNRAS.493L.120K}
}

@ARTICLE{Lu2023,
       author = {{Lu}, Wenbin and {Quataert}, Eliot},
        title = "{Quasi-periodic eruptions from mildly eccentric unstable mass transfer in galactic nuclei}",
      journal = {\mnras},
         year = 2023,
        month = oct,
       volume = {524},
       number = {4},
        pages = {6247-6266},
          doi = {10.1093/mnras/stad2203},
archivePrefix = {arXiv},
       eprint = {2210.08023},
 primaryClass = {astro-ph.HE},
       adsurl = {https://ui.adsabs.harvard.edu/abs/2023MNRAS.524.6247L}
}

@ARTICLE{Xian2021,
       author = {{Xian}, Jingtao and {Zhang}, Fupeng and {Dou}, Liming and {He}, Jiasheng and {Shu}, Xinwen},
        title = "{X-Ray Quasi-periodic Eruptions Driven by Star-Disk Collisions: Application to GSN069 and Probing the Spin of Massive Black Holes}",
      journal = {\apjl},
         year = 2021,
        month = nov,
       volume = {921},
       number = {2},
          eid = {L32},
        pages = {L32},
          doi = {10.3847/2041-8213/ac31aa},
archivePrefix = {arXiv},
       eprint = {2110.10855},
 primaryClass = {astro-ph.HE},
       adsurl = {https://ui.adsabs.harvard.edu/abs/2021ApJ...921L..32X}
}

@ARTICLE{Tagawa2023,
       author = {{Tagawa}, Hiromichi and {Haiman}, Zolt{\'a}n},
        title = "{Flares from stars crossing active galactic nucleus discs on low-inclination orbits}",
      journal = {\mnras},
         year = 2023,
        month = nov,
       volume = {526},
       number = {1},
        pages = {69-79},
          doi = {10.1093/mnras/stad2616},
archivePrefix = {arXiv},
       eprint = {2304.03670},
 primaryClass = {astro-ph.HE},
       adsurl = {https://ui.adsabs.harvard.edu/abs/2023MNRAS.526...69T}
}

@ARTICLE{Franchini2023,
       author = {{Franchini}, Alessia and {Bonetti}, Matteo and {Lupi}, Alessandro and {Miniutti}, Giovanni and {Bortolas}, Elisa and {Giustini}, Margherita and {Dotti}, Massimo and {Sesana}, Alberto and {Arcodia}, Riccardo and {Ryu}, Taeho},
        title = "{Quasi-periodic eruptions from impacts between the secondary and a rigidly precessing accretion disc in an extreme mass-ratio inspiral system}",
      journal = {\aap},
         year = 2023,
        month = jul,
       volume = {675},
          eid = {A100},
        pages = {A100},
          doi = {10.1051/0004-6361/202346565},
archivePrefix = {arXiv},
       eprint = {2304.00775},
 primaryClass = {astro-ph.HE},
       adsurl = {https://ui.adsabs.harvard.edu/abs/2023A&A...675A.100F}
}

@ARTICLE{Linial2023b,
       author = {{Linial}, Itai and {Metzger}, Brian D.},
        title = "{EMRI + TDE = QPE: Periodic X-Ray Flares from Star-Disk Collisions in Galactic Nuclei}",
      journal = {\apj},
         year = 2023,
        month = nov,
       volume = {957},
       number = {1},
          eid = {34},
        pages = {34},
          doi = {10.3847/1538-4357/acf65b},
archivePrefix = {arXiv},
       eprint = {2303.16231},
 primaryClass = {astro-ph.HE},
       adsurl = {https://ui.adsabs.harvard.edu/abs/2023ApJ...957...34L}
}

@ARTICLE{Zhou2024a,
       author = {{Zhou}, Cong and {Huang}, Lei and {Guo}, Kangrou and {Li}, Ya-Ping and {Pan}, Zhen},
        title = "{Probing orbits of stellar mass objects deep in galactic nuclei with quasiperiodic eruptions}",
      journal = {\prd},
         year = 2024,
        month = may,
       volume = {109},
       number = {10},
          eid = {103031},
        pages = {103031},
          doi = {10.1103/PhysRevD.109.103031},
archivePrefix = {arXiv},
       eprint = {2401.11190},
 primaryClass = {astro-ph.HE},
       adsurl = {https://ui.adsabs.harvard.edu/abs/2024PhRvD.109j3031Z}
}

@ARTICLE{Jiang2025,
       author = {{Jiang}, Ning and {Pan}, Zhen},
        title = "{Embers of Active Galactic Nuclei: Tidal Disruption Events and Quasiperiodic Eruptions}",
      journal = {\apjl},
         year = 2025,
        month = apr,
       volume = {983},
       number = {1},
          eid = {L18},
        pages = {L18},
          doi = {10.3847/2041-8213/adc456},
archivePrefix = {arXiv},
       eprint = {2503.17609},
 primaryClass = {astro-ph.HE},
       adsurl = {https://ui.adsabs.harvard.edu/abs/2025ApJ...983L..18J}
}

@INPROCEEDINGS{Kaastra1996,
       author = {{Kaastra}, J.~S. and {Mewe}, R. and {Nieuwenhuijzen}, H.},
        title = "{SPEX: a new code for spectral analysis of X \& UV spectra.}",
    booktitle = {UV and X-ray Spectroscopy of Astrophysical and Laboratory Plasmas},
         year = 1996,
       editor = {{Yamashita}, K. and {Watanabe}, T.},
        month = jan,
        pages = {411-414},
       adsurl = {https://ui.adsabs.harvard.edu/abs/1996uxsa.conf..411K}
}

\end{document}